\documentstyle{article}
\begin{document}

\newcommand{\coment}[1]{ }

\thispagestyle{empty}

\begin{center}
\large
M. V. Lomonosov Moscow State University \\
D. V. Scobeltsyn Institute of Nuclear Physics \\
\vskip 2cm
\end{center}
\begin{flushright}
INP MSU 96--24/431
\end{flushright}
\vskip 2cm
\begin{center}
{\LARGE \bf L~a~n~H~E~P ---} \\[3mm]
{\Large a package for automatic generation \\
of Feynman rules in gauge models} \\[1.5cm]
\vskip 1.5cm
{\Large A. V. Semenov}
\end{center}
\vfill
\begin{center}
Moscow \\
{ \it 1996 }
\end{center}

\eject

\thispagestyle{empty}
{\Large LanHEP --- a package for automatic generation of Feynman rules
in gauge models.}

\vfill

{\bf Abstract}.  The LanHEP program for Feynman
rules generation in momentum representation is presented. It reads
the Lagrangian written in the compact form close to one used in 
publications. It means that Lagrangian terms can be written 
with summation over indices of broken symmetries and using special symbols
for complicated expressions, such as covariant derivative and  
strength tensor for gauge fields. The output
is Feynman rules in terms of physical fields and independent parameters.
This output can be written in LaTeX format and in the form of CompHEP
model files, which allows one to start calculations of processes in
the new physical model. Although this job is rather
straightforward and can
be done manually, it requires careful calculations and in the modern
theories with many particles and vertices can lead to errors and
misprints.
The program allows one to introduce into CompHEP new gauge theories as well
as various anomalous terms.   

\vskip 1cm
\begin{center}
\begin{tabular}{rl}
{\it E-mail:} & {\tt semenov@theory.npi.msu.su}\\
{\it WWW page:} & { \tt http://theory.npi.msu.su/\~{}semenov/lanhep.html}
\end{tabular}
\end{center}

\vfill
\begin{flushright} \copyright \ \ M. V. 
Lomonosov Moscow State University \\
D. V. Scobeltsyn Institute of Nuclear Physics 
\end{flushright}

\eject

\section*{Introduction}

LanHEP has been designed as a part of the CompHEP package \cite{chep},
 worked out for automatic calculations
in high energy physics. CompHEP allows symbolic computation
of the matrix element squared of any process with up to 6 incoming and
outgoing particles for a given physical model (i.e. a model defined by a
set of Feynman rules as 
a table of vertices in the momentum representation) and then numerical 
calculation of cross-sections and various distributions.

Main purpose of the new option given by LanHEP program is designing of
a new physical model. LanHEP program makes possible the generation of
Feynman rules for propagators and vertices in momentum representation
starting from the Lagrangian defined by a user in some simple format very
similar to canonical coordinate representation.
 User should prepare a text file with description of all Lagrangian terms
 in the format close to the form used in standard publications.
Of course, user has to describe all particles and
parameters appearing in Lagrangian terms.  

The main LanHEP
features are:
\begin{itemize}
\item LanHEP expands expression and combines similar terms;
\item it performs the Fourier transformation by replacing derivatives 
with momenta of particles; 
\item it writes Feynman rules in the form of four tables in CompHEP format
as well as tables in LaTeX format;
\item user can define the substitution rules, for example for
 covariant derivative;
\item it is possible to define multiplets, and (if necessary)
their components;
\item user can write Lagrangian terms with Lorentz and multiplet 
indices explicitly or omit indices (all or some of them);
\item LanHEP performs explicit summation over the indices in Lagrangian
terms, if the corresponding components for multiplets and matrices are
 introduced;
\item it allows the user to introduce vertices with 4 fermions and
4 colored particles (such vertices can't be introduced directly in
 CompHEP) by means of auxiliary field with constant propagator;
\item it also can check whether the set of introduced vertices satisfies
 the electric charge conservation law.
\end{itemize}

\section{QED by means of LanHEP}

We start from a simple exercises, illustrating the
 main ideas and features of LanHEP. The first physical model is
Quantum Electrodynamics.

\begin{figure}[h]
\framebox{ \vbox{
\flushleft \tt \hspace*{1cm} model QED/1. \\
\hspace*{1cm} parameter ee=0.31333:'elementary electric charge'.\\
\hspace*{1cm} spinor e1/E1:(electron, mass me=0.000511). \\
\hspace*{1cm} vector A/A:(photon). \\
\hspace*{1cm} let F\^{}mu\^{}nu=deriv\^{}nu*A\^{}mu-deriv\^{}mu*A\^{}nu. \\
\hspace*{1cm} lterm -1/4*(F\^{}mu\^{}nu)**2 - 
		1/2*(deriv\^{}mu*A\^{}mu)**2. \\
\hspace*{1cm} lterm E1*(i*gamma*deriv+me)*e1. \\
\hspace*{1cm} lterm ee*E1*gamma*A*e1. }
}
\caption{LanHEP input file for the generation of QED Feynman rules}
\label{fig:qed1}
\end{figure}

QED  Lagrangian is
$$ {\cal L}_{QED}=-\frac{1}{4}F_{\mu\nu} F^{\mu\nu} +
\bar e \gamma^\mu(i\partial_\mu +
g_e A_\mu)e - m\bar e e   $$
and the gauge fixing term in Feynman gauge has the form
$$ {\cal L}_{GF}=-\frac{1}{2}(\partial_\mu A^\mu)^2.$$
Here $e(x)$ is the spinor electron-positron field, $m$ is
 the electron mass, 
$A_\mu(x)$ is the vector photon field, 
$F^{\mu\nu}=\partial_\nu A^\mu-\partial_\mu A^\nu$,
  and $g_e$ is the elementary electric charge.

The LanHEP input file to generate the Feynman rules for QED is 
shown in Fig. \ref{fig:qed1}.

First of all, the input file consists of
statements. Each statement begins with one of the reserved keywords
 and ends
 by the full-stop '.' symbol.

First line says that this is a model with the name {\tt QED} and number 1.
This information is supplied for CompHEP, the name {\tt QED} will be 
displayed in its list of models.
In CompHEP package each model is described by four files:
 'varsN.mdl', 'funcN.mdl', 'prtclsN.mdl', 'lgrngnN.mdl' , where N is the
very number specified in the {\tt model} statement.

The {\tt model} statement stands first in the input file. If this statement
is absent, LanHEP does not generate four standard CompHEP files, just
 builds the model and prints
 diagnostic if  errors are found.

Second line in the input file contains declaration of the model parameter,
denoting elementary electric charge $g_e$ as {\tt ee}. For each parameter
 used in the model one should declare its
numeric value and optional comment (it is also used in CompHEP menus).

The next two lines declare particles.
Statement names {\tt spinor, vector}
correspond to the particle spin. So, we declare electron 
 denoted by
{\tt e1} (the corresponding antiparticle name is {\tt E1}) and
photon denoted by {\tt A} (with antiparticle
name being  {\tt A}, since the antiparticle for photon is identical
to particle).

After particle name we give in
brackets some
options. The first one is full name of the particle, used in
CompHEP; the second option declares the
 mass of this particle.

The {\tt let} statement in the next  line declares the substitution rule
for symbol
{\tt F}, which will be replaced in the further Lagrangian terms
 by the expression given in this statement.

Predefinite name {\tt deriv}, reserved for the derivation $\frac{\partial}
{\partial  x}$, will be replaced after 
the Fourier transformation by the momentum of the particle multiplied
 by $-i$.

The rest of the lines describe Lagrangian terms. Here the reserved name
{\tt gamma} denotes Dirac's $\gamma$-matrices.

One can see that the indices  are written separated with the
caret symbol '\^{}'.
Note that in the last two lines we have omitted indices. It means that
LanHEP  restores omitted indices automatically. Really, one can type
the last term in the full format: \begin{quote}
{\tt lterm ee*E1\^{}a*gamma\^{}a\^{}b\^{}mu*A\^{}mu*e1\^{}b}. \end{quote}
It corresponds to $g_e \bar e_a \gamma^\mu_{ab} e_b A_\mu$ 
with all indices written. Note that the order of objects
 in the monomial is important to restore indices automatically.

\section{QCD}

Now let us consider the case of the  Quantum
Chromodynamics. The Lagrangian for gluon fields reads

$$ L_{YM} = -\frac{1}{4}F^{a\mu\nu}F^a_{\mu\nu},$$
where
$$
F^a_{\mu\nu}=\partial_\mu G^a_\nu-\partial_\nu G^a_\mu-
	g_s f^{abc}G^b_\mu G^c_\nu,$$
$G^a_\mu(x)$ is the gluon field, $g_s$ is a strong charge
 and $f^{abc}$ are purely imaginary structure
constants of $SU(3)$ color group.

The quark kinetic term and its interaction  with the gluon has the form
$$ L_F = \bar q_i \gamma^\mu \partial_\mu q_i + g_s  \lambda^a_{ij} 
\bar q_i\gamma^\mu  q_j G_\mu^c,$$
where $\lambda^a_{ij}$ are Gell-Mann matrices.

Gauge fixing terms in Feynman gauge together with the corresponding 
Faddev-Popov ghost term  are
$$ -\frac{1}{2}(\partial_\mu G^\mu_a)^2 + ig_s f^{abc} \bar c^a G^b_\mu
\partial^\mu c^c,$$
where $(c, \bar c)$ are unphysical ghost fields.

The corresponding LanHEP input file is shown in Fig. \ref{fig:qcd1}.

\begin{figure}[th]
\framebox{ \vbox{
\flushleft \tt \hspace*{1cm} model QCD/2. \\
\hspace*{1cm} parameter gg=1.117:'Strong coupling'.\\
\hspace*{1cm} spinor q/Q:(quark, mass mq=0.01, color c3). \\
\hspace*{1cm} vector G/G:(gluon, color c8, gauge). \\
\hspace*{1cm} let F\^{}mu\^{}nu\^{}a = deriv\^{}nu*G\^{}mu\^{}a -
           deriv\^{}mu*G\^{}nu\^{}a -
\hspace*{2cm} gg*f\_SU3\^{}a\^{}b\^{}c*G\^{}mu\^{}b*G\^{}nu\^{}c.\\
\hspace*{1cm} lterm -F**2/4-(deriv*G)**2/2.\\
\hspace*{1cm} lterm Q*(i*gamma*deriv+mq)*q.\\
\hspace*{1cm} lterm i*gg*f\_SU3*ccghost(G)*G*deriv*ghost(G). \\
\hspace*{1cm} lterm gg*Q*gamma*lambda*G*q. }
}
\caption{Input file for the generation of QCD Feynman rules}
\label{fig:qcd1}
\end{figure}

\begin{table}[ht]
\caption{QCD Feynman rules generated by LanHEP in LaTeX output format}
\begin{center}
\begin{tabular}{|llll|l|} \hline
\multicolumn{4}{|c|}{Fields in the vertex} & Variational derivative
 of Lagrangian by fields \\ \hline
${G}_{\mu p }$ & ${G.C}_{q }$ & ${G.c}_{r }$ &  & $- gg\cdot p_3^\mu
 f_{p q r} $\\[2mm]
${Q}_{a p }$ & ${q}_{b q }$ & ${G}_{\mu r }$ &  & $ gg\cdot
 \gamma_{a b}^\mu \lambda_{p q}^r $\\[2mm]
${G}_{\mu p }$ & ${G}_{\nu q }$ & ${G}_{\rho r }$ &  & $ ggf_{p q r}
 \big(p_3^\nu g^{\mu \rho} -p_2^\rho g^{\mu \nu} -p_3^\mu g^{\nu \rho}
 +p_1^\rho g^{\mu \nu} +p_2^\mu g^{\nu \rho} -p_1^\nu g^{\mu \rho}
 \big)$\\[2mm]
${G}_{\mu p }$ & ${G}_{\nu q }$ & ${G}_{\rho r }$ & ${G}_{\sigma s }$
 & $ gg^2 \big(g^{\mu \rho} g^{\nu \sigma} f_{p q t} f_{r s t} -g^{\mu
 \sigma} g^{\nu \rho} f_{p q t} f_{r s t} +g^{\mu \nu} g^{\rho \sigma}
 f_{p r t} f_{q s t} $ \\[2mm]
 & & & & $+g^{\mu \nu} g^{\rho \sigma} f_{p s t} f_{q r t} -g^{\mu
 \sigma} g^{\nu \rho} f_{p r t} f_{q s t} -g^{\mu \rho} g^{\nu \sigma}
 f_{p s t} f_{q r t} \big)$\\ \hline
\end{tabular}
\end{center}
\end{table}

Since QCD uses objects with
color indices, one has to declare the indices of these objects.
There are three types of color indices supported by LanHEP. These types 
are referred as {\tt color c3} (color triplets), {\tt color c3b}
 (color antitriplets), and {\tt color c8} (color octets). 
One can see that {\tt color c3} index type appears among the options in
the quark {\tt q} declaration, and the {\tt color c8} one in the gluon
{\tt G} declaration. Antiquark {\tt Q} has got color index of type {\tt 
color c3b} as antiparticle to quark.
 LanHEP allows  convolution of an index of type {\tt color c3} only with
another index of type {\tt color c3b}, and two indices of type {\tt color
c8}. Of course, in Lagrangian terms each index
has to be convoluted with its partner, since Lagrangian has to be scalar.

LanHEP allows also to use in the Lagrangian terms a predefined symbol 
{\tt lambda} with the three indices of types {\tt color c3, color c3b,
color c8} corresponding to Gell-Mann  
$\lambda$-matrices. Symbol {\tt f\_SU3} denotes the structure constant
$f^{abc}$ of color $SU(3)$
group (all three indices have the type {\tt color c8}).

Option {\tt gauge}
in the declaration of {\tt G} allows to use names {\tt ghost(G)} and {\tt
ccghost(G)} for the ghost fields $c$ and $\bar c$ in Lagrangian terms
and in {\tt let} statements. 

Table 1 shows Feynman rules generated by LanHEP in LaTeX format 
after processing the input file presented in Fig. \ref{fig:qcd1}. 
Four gluon  vertex rule is indicated in the last line. Note that the
output in CompHEP format has no 4-gluon vertex explicitly; it is 
expressed effectively through 3-leg vertices by the constant propagator
 of some auxiliary field (see Appendix \ref{imp} for more details).


\section{Syntax}

The LanHEP input file is the sequence of statements,
each starts with a special identifier (such as {\tt parameter,
lterm} etc) and ends with the full-stop '{\bf .}' symbol. 
Statement can occupy several lines in the input file.

This section is aimed to clarify the syntax of LanHEP input files, i.e. 
the structure of the statements.

\subsection{Constants and identifiers}

First of all, each word in any statement is either an {\it identifier}
or a {\it constant}.

 Indentfiers are the names of particles, parameters etc.
Examples of identifiers from the previous section are particle names 
\begin{quote} {\tt e1 E1 A q Q G} \end{quote}
 The first word in each statement is also
an identifier, defining the function which this statement
performs. The identifiers are usually combinations of letters and digits
starting with a letter. If an identifier doesn't respect this 
rule, it should be quoted. For example, the names of $W^\pm$ bosons
must be written as {\tt 'W+'} and {\tt 'W-'}, since they contain 
'+' and '--' symbols.

Constants can be classified  as
\begin{itemize}
\item {\it integers}: they consist of optional sign followed by one or
more decimal digits, such as  
\begin{quote} { \tt 0 \, 1 \, -1 \, 123 \, -98765} \end{quote}
Integers can  appear in Lagrangian terms, parameter definition and
in other expressions.
\item {\it Floating point numbers} 
 include optional sign, several decimal digits of
mantissa with an embedded period (decimal point) with at least one digit
before and after the period, and optional exponent. The exponent, if
present, consist of letter {\tt E} or {\tt e} followed by an optional sign
and one or more decimal digits. The valid examples of floating point
 numbers are 
\begin{quote} {\tt 1.0 \, -1.0 \, 0.000511 \, 5.11e-4} \end{quote}
Floating point numbers are used only as parameter values (coupling
constants, particle masses etc). They can not
be explicitly used in Lagrangian terms.
\item {\it String constants} may include arbitrary symbols. They are
used as comments in parameter statements, full particle
names in the declaration of a particle, etc. Examples from the previous
section are 
\begin{quote} {\tt electron \\ photon} \end{quote} 
If a string constant
contains any character besides letters and digits or doesn't begin 
with a letter, it should be quoted. For example, the comments in QED and
 QCD input files (see previous section)
 contain blank spaces, so they are quoted:
\begin{quote} {\tt 'elementary electric charge' \\ 'Strong coupling'}
\end{quote}
\end{itemize}

\subsection{Comments}

User can include comments into the LanHEP input file in two ways. First,
symbol {\tt '\%'} denotes the comment till the end of current line.
Second way allows one to comment any number of lines by putting a part of
input file  between
{\tt '/*'} (begin of comment) and {\tt '*/'} (end of comment) symbols.


\section{Objects in the expressions for Lagrangian terms}

Each symbol which may appear in algebraic expressions (names of
parameters, fields, etc) has a fixed order of indices and their types. 
If this object is used in any expression, one should write its indices in
 the same order as they were defined when the object has been declared.

Besides the indices types corresponding to color $SU(3)$ group: {\tt color
 c3}
 (color triplet), {\tt color
c3b} (color antitriplet) and {\tt color c8} (color octet)
described in the previous example, there are default types 
of indices for Lorentz group:  {\tt vector}, {\tt spinor} and {\tt cspinor}
(antispinor). User can also declare
new types of indices corresponding to the 
symmetries other than color $SU(3)$ group. In this case any object
(say, particle)
may have indices related to this new group. This possibility will be
described in Section \ref{newgrp}.

If an index appears twice in some monomial of an expression, LanHEP
assumes summation over this index. Types of such indices must allow
the convolution, i.e. they should be one of the pairs:
{\tt spinor} and {\tt cspinor}, two {\tt vector}, {\tt color c3} and
{\tt color c3b}, two {\tt color c8}.  

In general the following objects are available to appear in the expressions
for a Lagrangian: integers and identifiers of parameters, particles,
 specials,
let-substitutions and arrays.

There are also predefined symbols {\tt i}, denoting imaginary unit $i$
($i^2=-1$) and {\tt Sqrt2}, which is a parameter with value the
$\sqrt{2}$.

\subsection{Parameters}

{\it Parameters\/} are scalar objects (i.e. they have no indices).
Parameters denote coupling constants,
masses and widths of particles, etc. To introduce a new parameter 
 one should use the  {\tt parameter} statement,
which  has the generic form
\begin{quote} { \tt parameter \sl name\/\tt =\sl value\/\tt :\sl comment. }
\end{quote}
\begin{itemize}
\item \underline{\sl name} is an identifier of newly created parameter. 
\item \underline{\sl value} is an integer or floating point number or an
expression. One can use previously declared parameters
and integers joined by standard arithmetical operators {\tt '+', '-',
 '*', '/'}, and {\tt '**'} (power).  
\item \underline{\sl comment }
is an optional comment to clarify the meaning of parameter,
it is used in CompHEP help windows. 
Comment has to be a string constant, so if it contains blank spaces
or other special characters, it must be quoted (see Section 3). 
\end{itemize}

\subsection{Particles}

{\it Particles\/ } are objects to denote physical particles. 
They may have indices. It is possible to use three statements to declare
a new particle, at the same time the second and the third statements define
the corresponding Lorentz index:
 \begin{quote}
{ \tt  scalar \sl P/aP\tt :(\sl options\tt ).} \\
{ \tt  spinor \sl P/aP\tt :(\sl options\tt ). } \\
{ \tt  vector \sl P/aP\tt :(\sl options\tt ). }   \end{quote}
{\sl P} and {\sl aP} are identifiers of particle and antiparticle.
In the case of truly neutral particles (when antiparticle is identical to 
the particle itself) one should use the form {\sl P/P} with identical
 names for particle and antiparticle. 

It is possible to write only the particle name, e.g. 
\begin{quote} {\tt  scalar  \sl P\tt :(\sl options\tt )}. \end{quote}
 In this case the name of corresponding
antiparticle is generated automatically. It satisfies the usual CompHEP
convention, when the name of antiparticle differs from particle by 
altering the case of the first letter. 
So for electron name {\tt e1} automatically generated antiparticle
name will be {\tt E1}. If the name contains symbol '+' it is replaced by
'--' and vice versa.

The {\sl option} is comma-separated list of options for a declared
 particle,  and it may include the following items:
\begin{itemize}
\item the first element in this list must be the full name of the particle,
   (e.g. {\tt electron} and {\tt photon} in our example.) 
      Full name is string constant, so it 
        should be quoted if it contains blanks, etc.
\item \underline{\tt mass \sl param=value} defines the mass of the
particle.
Here {\sl param} is an identifier of new parameter, which is used to
denote the mass; {\sl  value} is its value,  
 it has the same syntax as in the {\tt parameter} statement, comment for
 this new parameter being generated automatically. If this option omitted,
 the mass is assumed to be zero.
\item \underline{\tt width \sl param=value} declares the width of the
 particle. It has the same syntax as for {\tt mass } option.
\item \underline{\sl itype} is a type of index of some symmetry; 
one can use default index types for color $SU(3)$ group (see QCD 
example in Section 2). It is possible to use user-defined index types
 (see Section \ref{newgrp}) and Lorentz group indices {\tt vector,
 spinor, cspinor}. 
\item \underline{\tt left} or \underline{\tt right} say that the
massless spinor particle is an eigenstate of $(1-\gamma^5)/2$ or
 $(1+\gamma^5)/2$ projectors, so this fermion is left-handed or
right-handed one.
\item \underline{\tt gauge} declares the vector particle as a gauge boson.
 This  option generates corresponding ghosts and goldstone bosons names
 for the named particle (see below).
\end{itemize}

When a particle name is used in any expression (in Lagrangian terms), one
should remember that the first index is either vector or spinor
one (of course, if this particle is not a Lorentz scalar). Then the
 indices follow in the
same order as index types in the {\sl options} list. So, in the case of 
quark declaration (see QCD example) the first index is spinor, and the
second one is color triplet. 

There are several functions taking particle name as an argument which
can be used in algebraic expressions. These functions are replaced with 
auxiliary particle names, which are generated automatically.
\begin{itemize}
\item Ghost field names in gauge theories are generated by the functions
 {\tt ghost(\sl name\tt) $\rightarrow$ '\sl name\/\tt .c'} and
 {\tt ccghost(\sl name\tt) $\rightarrow$ '\sl name\/\tt .C'} (see for
 instance
 Table 1).
 Here and below
{\sl name} is the name of the corresponding gauge boson. 
\item Goldstone boson field name in the t'Hooft-Feynman gauge is generated
 by
the function  {\tt gsb(\sl name\tt) $\rightarrow$ '\sl name\/\tt .f'}.
\item The function {\tt anti(\sl name\/\tt)} generates antiparticle name
for the  particle {\sl name}. 
\item The name for a charge conjugated spinor particle 
$\psi^c=C\bar{\psi}{}^T$ is generated by the function 
 {\tt cc(\sl name\/\tt) $\rightarrow$ '\sl name\/\tt .c'}. Charge 
conjugated fermion has the same indices types and ordering, however index
of  {\tt spinor} type is replaced by the index type {\tt cspinor} and
 vice versa.
\item {\tt vev(\sl expr\tt\/)} is used in Lagrangian terms for vacuum
 expectation values.
Function {\tt vev} ensures that {\tt deriv*vev(\sl expr\/\tt)} is zero.
In other words, {\tt vev} function forces LanHEP to treat {\sl expr} as
 a scalar particle
which will be replaced by {\sl expr} in Feynman rules.
\end{itemize}

\subsection{Specials}

Besides parameters and particles other indexed
such as $\gamma$-matrices, group structure constants, etc may appear in the
Lagrangian terms. 
We refer such objects as {\it specials}. 

Predefined specials of Lorentz group are:
\begin{itemize}
\item \underline{\tt gamma} stands for the $\gamma$-matrices. It has three
    indices of  {\tt spinor, cspinor} and {\tt vector} types.
\item \underline{\tt gamma5} denotes $\gamma^5$ matrix. It has two
    indices of  {\tt spinor} and {\tt cspinor}  types.
\item \underline{\tt moment} has one index of {\tt vector} type.
At the stage of Feynman rules generation this symbol is replaced by the
 particle moment.
\item \underline{\tt deriv}  is replaced 
by $-ip_\mu$, where $p_\mu$ is the particle moment. It has one {\tt vector}
 index.
\end{itemize}

Specials of color $SU(3)$ group are:
\begin{itemize}
\item \underline{\tt lambda} denotes Gell-Mann $\lambda$-matrices. It has
 three
indices: {\tt color c3, color c3b} and {\tt color c8}.
\item \underline{\tt f\_SU3} is the $SU(3)$ structure constant. It has
 three indices
of {\tt color c8} type.
\end{itemize}

Note that for specials the order of indices types is fixed. 
  
User can  declare new specials with the help of a facility to introduce
 user-defined
indices types (see Section~\ref{newgrp}). 

\subsection{Let-substitutions}

LanHEP allows the user to introduce new symbols and then substitute them
in Lagrangian terms by some
expressions. 
Substitution has the generic
form \begin{quote} {\tt let \sl name\/\tt =\sl expr. }\end{quote}
where {\sl name} is the identifier of newly defined object.
 The expression has the same structure as those in Lagrangian
terms, however here expression may have free (non-convoluted) indices.

Typical example of using a
substitution rule is a definition of the QED  covariant derivative as
\begin{quote} {\tt let Deriv\^{}mu=deriv\^{}mu + i*ee*A\^{}mu. }\end{quote}
corresponding to $D_\mu=\partial_\mu + ig_eA_\mu$.

More complicated example is the declaration $\sigma^{\mu\nu}
\equiv i(\gamma^\mu\gamma^\nu-\gamma^\nu\gamma^\mu)/2$ matrices:
\begin{quote} 
{\tt let sigma\^{}a\^{}b\^{}mu\^{}nu = 
i*(gamma\^{}a\^{}c\^{}mu*gamma\^{}c\^{}b\^{}nu 
- gamma\^{}a\^{}c\^{}nu*gamma\^{}c\^{}b\^{}mu)/2.}
\end{quote}

Note that the order of indices types 
of new symbol is fixed by the declaration. So, first two indices of
{\tt sigma} after this declaration
are spinor and antispinor, third and fourth are vector indices.

\subsection{Arrays}

LanHEP allows to define components of indexed objects. In this case,
convolution of indices will be performed as an explicit sum of products
of the corresponding components.

Object with explicit components  has to be written as
\begin{quote} {\tt \{\sl expr1, expr2 ..., exprN \tt \}\^{}\sl i
 }\end{quote}
where  expressions correspond to components. All indices of components
(if they present) have to be written at each component, and the index 
numbering components  has to be written after closing curly bracket. 
Of course, all the components must have the same types of free
(non-convolved) indices.  

Arrays are usually applied for the definition of multiplets and matrices
corresponding to broken symmetries. 

Typical example of arrays usage is a declaration of
electron-neutrino isospin doublet {\tt l1} (and antidoublet {\tt L1})
\begin{quote} {\tt let l1\^{}a\^{}I = \{ n1\^{}a, e1\^{}a\}\^{}I, 
L1\^{}a\^{}I = \{ N1\^{}a, E1\^{}a\}\^{}I. } \end{quote}
Here we suppose that {\tt n1} was declared as the spinor particle
 (neutrino), with the antiparticle name {\tt N1}.

Matrices can be represented as arrays which have other arrays as 
components.
However, it is more convenient to declare them with omitted indices, see
Section \ref{oimatr} (the same is correct for multiplets also).
 
It is possible also to use arrays directly in the Lagrangian terms, rather
 than  only in the declaration of let-substitution.


\section{Structure of expressions for the Lagrangian}

When all parameters and particles necessary for introduction of
physical model are declared, one can enter Lagrangian terms with the help
of  the 
 {\tt lterm} statement:
\begin{quote} {\tt lterm \sl expr\tt . }\end{quote}

Elementary objects of expression are integers, identifiers of parameters,
particles, specials, let-substitutions, and arrays.

These elementary objects can be combined by usual arithmetical operators
as 
\begin{itemize}
\item  {\sl expr1+expr2\/}  (addition),
\item  {\sl expr1--expr2\/}  (subtraction),
\item {\sl expr1*expr2\/}  (product),
\item {\sl expr1/expr2\/}  (fraction; here {\sl expr2} must be a product
 of integers and parameters),
\item {\sl expr1**N\/}  ({\sl N}th power of {\sl expr1}; {\sl N}
 must be integer).
\end{itemize}
  One can use brackets '(' and ')' to force  the precedence
of operators. Note, that indices can follow only elementary objects
symbols, i.e. if {\tt A1} and {\tt A2} were declared as two vector
 particles then valid
expression for their sum is {\tt A1\^{}mu+A2\^{}mu}, rather than {\tt
(A1+A2)\^{}mu}.

\subsection{Where-substitutions}

More general form of expressions  involves {\it where-substitutions}:
\begin{quote} {\sl expr \tt where \sl subst}.\end{quote}
 
 In the simple form {\sl subst} is
 {\sl name\tt =\sl repl} or several constructions of such kind separated by
comma ','. In the form of such kind each instance of identifier {\sl name}
 in {\sl expr} is
replaced by {\sl repl}. 

Note that in contrast to let-substitutions,
where-substitution doesn't create a new object. LanHEP simply replaces
 {\sl name} by 
{\sl repl}, and then processes the resulting expression. It means in
 particular
that  {\sl name} can not have indices, although it can denote an object
 with
indices:
\begin{quote} {\tt lterm F**2 
where F=deriv\^{}mu*A\^{}nu-deriv\^{}nu*A\^{}mu.}
\end{quote}
is equivalent to 
\begin{quote} {\tt lterm (deriv\^{}mu*A\^{}nu-deriv\^{}nu*A\^{}mu)**2.}
\end{quote} 

 The substitution rule introduced by 
the keyword {\tt where} is active only within the current {\tt lterm}
 statement.  

More general form of where-substitution allows to use several
{\sl name\/\tt =\sl repl} 
substitution rules separated by semicolon ';'. In this
case {\sl expr} will be replaced by the sum of expressions; each term
 in this sum
is produced by applying one of the substitution rules from
semicolon-separated list to the expression {\sl expr}.  This form is
 useful for writing
the Lagrangian where many particles have similar interaction. 

For example, if {\tt u,d,s,c,b,t} are declared as quark names,
their interaction with the gluon may read as
\begin{quote} {\tt lterm gg*anti(psi)*gamma*lambda*G*psi where \\
\phantom{lterm gg*} psi=u; psi=d; psi=s; psi=c; psi=b; psi=t.} \end{quote}

The equivalent form is
\begin{quote} {\tt lterm  gg*U*gamma*lambda*G*u + gg*D*gamma*lambda*G*d +
 \\
\phantom{lterm} gg*U*gamma*lambda*G*u + gg*D*gamma*lambda*G*d + \\
\phantom{lterm} gg*U*gamma*lambda*G*u + gg*D*gamma*lambda*G*d. }
 \end{quote}

Where-substitution can also be used in {\tt let} statement. In this case
one should use brackets:
\begin{quote} {\tt let \sl lsub=(expr\/\ \tt where \sl wsub=expr1).}
\end{quote} 

Note that in the previous example we have omitted indices. In the next
 section we shall describe this LanHEP option in details.


\section{Omitting indices}

Physicists usually do not write all possible indices in the Lagrangian
 terms.
LanHEP also allows a user
to omit indices. This feature can simplify introduction of
expressions and make them more readable.
Compare two possible forms:
\begin{quote} {\tt lterm E1\^{}a*gamma\^{}a\^{}b\^{}mu*A\^{}mu*e1\^{}b. }
 \end{quote}
corresponding to $g_e \bar e_a(x) \gamma^\mu_{ab} e_b(x) A_\mu(x)$, and
\begin{quote} {\tt lterm E1*gamma\^{}mu*e1*A\^{}mu. }\end{quote}
corresponding to $g_e \bar e(x) \gamma^\mu e(x) A_\mu(x)$). 
Furthermore, while physicists usually write vector indices explicitly in 
the formulas, in LanHEP  vector indices also can be omitted:
\begin{quote} {\tt lterm -i*ee*E1*gamma*e1*A. }\end{quote}
Generally speaking, when user omits indices in the expressions, LanHEP
 faces two  problems: which
indices were omitted  and
how to convolute restored indices. 

\subsection{Restoring the omitted indices}

When the indexed object  is declared
 the corresponding set of indices is assumed.
 Thus, if the quark {\tt
q} is declared as
\begin{quote} {\tt spinor q:('some quark', color c3). }\end{quote}
its first index is spinor and the second one belongs to the {\tt color c3}
type.
If both indices are omitted in some expression, LanHEP generates
them in the correspondence to  order ({\tt spinor, color c3}).
However, if only one index is written, for example in
the form  {\tt q\^{}a}, LanHEP has to recognize
whether the index {\tt a} is of {\tt color c3} or of  {\tt spinor}
types.

To solve this problem LanHEP looks up the {\it list of indices
omitting order}.
By default this list is set to \begin{quote}
{\tt [spinor, color c3, color c8, vector]} \end{quote}

The algorithm to restore omitted indices is the following. First, LanHEP
assumes that user has omitted
indices which belong to the first type (and corresponding antitype) from
 this
list. Continuing the consideration of our example with particle {\tt q}
 one can
see that since this particle is declared having one {\tt spinor} index (the
first type  in the list) LanHEP checks whether the number of
indices declared for this object without {\tt spinor} index equals to the
number of indices written explicitly by user. In our example (when user
has written {\tt q\^{}a}) this is true. In the following LanHEP concludes 
that the user omitted {\tt spinor} index and that  explicitly written
 index is
of {\tt color c3} type.

In other cases, when the supposition fails if the user has omitted indices
 of the 
first type in the {\it list of indices omitting order}, LanHEP goes to
the second step. It assumes that user
has omitted indices of first two
types from this list. If this assumption also fails, LanHEP 
assumes that user has omitted indices of first three types in the list and
 so on. 
At each step LanHEP subtracts the number  of  indices of
these types assumed to be omitted from the full number of indices declared
 for the
object, and checks
whether this number of resting indices equals to the number of 
explicitly written indices.  If LanHEP fails when the {\it list of indices
omitting order} is completed, error message is returned by the program.

Note that if the user would like to omit indices of some type, he must 
omit 
all indices of this type (and antitype) as well as  the
indices of all types which precede in the {\it list of indices omitting
order}.

For example, if object {\tt Y} is declared  with one {\tt spinor},
two {\tt vector} and three {\tt color c8} indices, than 
\begin{itemize}
\item the form {\tt Y\^{}a\^{}b\^{}c\^{}d\^{}e\^{}f} means that the user
 wrote 
all the indices explicitly;
\item the form {\tt Y\^{}a\^{}b\^{}c\^{}d\^{}e} means that the user
 omitted {\tt spinor} 
index and wrote {\tt vector} and {\tt color c8} ones;
\item the form {\tt Y\^{}a\^{}b} means that the user omitted {\tt spinor}
 and {\tt color c8}
indices and wrote only two {\tt vector} ones;
\item the form {\tt Y} means that the user omitted all indices;
\item all other forms, involving different number of written indices, are
 incorrect.
\end{itemize}

One can say that indices should be omitted in the direct correspondence
with abrupting the {\it list of indices omitting order} from left to right.

One could change the {\it list of indices omitting order
types} by the statement {\tt SetDefIndex}. For example, for default
setting it looks like
\begin{quote} {\tt SetDefIndex(spinor, color c3, color c8, vector).
 }\end{quote}
Each argument in the list is a type of index.

\subsection{Convolution of restored indices}

Omitted index can be convoluted only with some another omitted index.
LanHEP expands the expression and restore indices in each monomial.
LanHEP reads objects in
the monomial  from the left to the right and checks whether
restored indices are present. If such index appears LanHEP seeks for the
 restored index of the appropriate type at the next objects. Note, that the
program does not check  whether the object with the first restored index
 has another
restored index of the appropriate type. Thus, if {\tt F} is declared 
as let-substitution for the
strength tensor of electromagnetic field (with two vector indices) then
expression {\tt F*F} (as well as {\tt F**2}) after processing omitted
 indices
turns to implied form {\tt \tt F\^{}mu\^{}nu*\tt F\^{}mu\^{}nu} rather
than {\tt \tt F\^{}mu\^{}mu*\tt F\^{}nu\^{}nu}.

This algorithm makes the convolutions to be sensitive to the order of
 objects in
the monomial. Let us look again to the QED example.
Expression {\tt E1*gamma*A*e1} (as well as {\tt A*e1*gamma*E1})
leads to correct result where vector index of photon is convoluted with the
same index of $\gamma$-matrix, spinor index of electron is convoluted with
antispinor index of $\gamma$-matrix and antispinor index of
positron is convoluted with spinor index of $\gamma$-matrix. However the
expression {\tt e1*E1*A*gamma} leads to wrong form {\tt
e1\^{}a*E1\^{}a*A\^{}mu*gamma\^{}b\^{}c\^{}mu}, because the first
 antispinor 
index after electron belongs to positron.
Spinor indices of {\tt gamma} stay free (non-convoluted) since no more
 objects with
appropriate indices (so, LanHEP will report an error since
Lagrangian term is not a scalar).

Note, that in the vertex with two $\gamma$-matrices the situation is more 
ambiguous. Let's look at  the
term corresponding to the electron anomalous magnetic moment
$\bar e(x) (\gamma^\mu\gamma^\nu-\gamma^\nu\gamma^\mu)e(x)F_{\mu\nu}$.
The correct LanHEP expression is 
\begin{quote} {\tt e1*(gamma\^{}mu*gamma\^{}nu
- gamma\^{}nu*gamma\^{}mu)*E1*F\^{}mu\^{}nu} \end{quote}
Here vector indices can't
be omitted, since it lead to the convolution of vector indices of 
$\gamma$-matrices. One can see also that the form 
\begin{quote} {\tt e1*E1*(gamma\^{}mu*gamma\^{}nu
- gamma\^{}nu*gamma\^{}mu)*F\^{}mu\^{}nu} \end{quote} will correspond to 
the expression $\bar e(x) e(x) Sp(\gamma^\mu\gamma^\nu)F_{\mu\nu}$. 
Here LanHEP has got scalar Lorentz-invariant expression in the Lagrangian
term, so it has no reason to report error. 

These examples mean that user
should clearly realize how the indices will be restored and
convoluted, or he has to write all indices explicitly.

\subsection{Let-substitutions}

Other problem arises when the omitted indices stay free, it is a case for
 the {\tt let} statement. 
LanHEP allows only two ways to avoid ambiguity in the order of indices
types:  either user
specifies all the indices at the name of new symbol and free indices in
 the corresponding
expression, or he should omit all free indices.
In the latter case the order of indices types is defined by the order of
free omitted indices in the first monomial of the expression. For
example
if {\tt A1} and {\tt A2} are vectors and {\tt c1} and {\tt c2} are spinors,
the statement \begin{quote} {\tt let d=A1*c1+c2*A2.} \end{quote}
declares new object {\tt d} with two indices, the first is vector index 
and the second  is spinor index according to their order in the monomial
{\tt A1*c1}. Of course, each monomial in the expression must have the same 
 typeset of free indices.

\subsection{Arrays \label{oimatr}}

The usage of arrays with omitted indices allows us to define matrices
conveniently.  For example, the declaration of
$\tau$-matrices 
\[
\tau_1=\left(\begin{array}{rr} 0 & 1 \\
 1 & 0 \end{array} \right), \;\;\;\;
\tau_2=\left(\begin{array}{rr} 0 & i \\ 
-i & 0 \end{array} \right), \;\;\;\;
\tau_3=\left(\begin{array}{rr} 1 & 0 \\ 0 & -1 \end{array} \right)
\]
can be written as
\begin{quote} {\tt
let tau1\ = \{\{0,\hphantom{-}1\}, \{1,\hphantom{-}0\}\}. \\
let tau2 = \{\{0,-i\}, \{i,\hphantom{-}0\}\}. \\
let tau3 = \{\{1,\hphantom{-}0\}, \{0,-1\}\}. } \end{quote}
One can see that in such way of declaration a matrix is written
"column by column".

The declaration of objects with three `explicit' indices can
 be done using the
objects already defined. For example, when $\tau$-matrices are defined as
before, it is easy
to define the vector  $\vec{\tau}\equiv(\tau_1, \tau_2, \tau_3)$ as
\begin{quote} {\tt let tau = \{tau1, tau2, tau3\}. } \end{quote}
The object {\tt tau} has three indices, first pair selects the element of
the matrix, while the matrix itself is selected by the third index, i.e.
{\tt tau\^{}i\^{}j\^{}a} corresponds to $\tau^a_{ij}$.

On the other hand, the declaration of structure constants of a group 
is more complicated.
Declaring such an object one should bear in mind that omitting indices
implies that in a sequence of components the second index of an object 
is changed after the full cycle of the first index, the third index is
 changed
after the full cycle of the second one, etc.
For example, a declaration of the antisymmetrical tensor
 $\varepsilon^{abc}$
can read as
\begin{quote} {\tt
          let eps =   \{\{\{0,0,0\}, \{0,0,-1\}, \{0,1,0\}\}, \\
\hphantom{let eps = } \{\{0,0,1\}, \{0,0,0\}, \{-1,0,0\}\}, \\
\hphantom{let eps = } \{\{0,-1,0\}, \{1,0,0\}, \{0,0,0\}\}\}.
  } \end{quote}
One can easily see that the components  are listed here in the
 following order:
\begin{quote}
$\varepsilon^{111}, \varepsilon^{211}, \varepsilon^{311},
 \varepsilon^{121},
\varepsilon^{221}, \varepsilon^{321}, \varepsilon^{131},
\; ... \;\varepsilon^{233}, \varepsilon^{333}$.\end{quote}
The declaration of more complex objects such as $SU(3)$ structure constants
can be made in the same way.

\section{Declaration of new index types and indexed objects \label{newgrp}}

\subsection{Declaring new groups}

Index type is defined by two  
keywords\footnote{The exception is Lorentz group, corresponding indices
types are defined by single keyword.}:   {\it group  name}
and   {\it representation name}. Thus, color triplet 
index type {\tt color c3} has group name {\tt color} and representation
name {\tt c3}.
 
LanHEP allows user to introduce new group names
by the  {\tt group} statement:
\begin{quote} {\tt group  \sl gname.}
\end{quote} Here {\sl gname } is a string constant,
 which becomes the name of newly declared group.

Representation names for each group name must be declared by the statement
\begin{quote} {\tt repres \sl gname\tt :(\sl rlist) }
\end{quote}
where  {\sl rlist } is a comma-separated list of representation names
declaration for the already declared group name {\sl gname}.
Each such declaration has the form either {\sl rname} or 
{\sl rname\tt /\sl crname}.  In the first case  index
which belongs to the {\sl gname rname} type can be convoluted with
another index of the same type; in the second case index of 
{\sl gname rname} type can be convoluted only with an index of 
{\sl gname crname} type.

For example, definition for color $SU(3)$ group with fundamental,
conjugated fundamental and adjoint representations looks as:
\begin{quote} { \tt group color:SU(3). \\ repres color:(c3/c3b,c8). }
\end{quote}
So, three index types can be used: {\tt color c3,
color c3b, color c8}. The convolution of these indices is allowed by
pairs ({\tt color c3, color c3b}) and ({\tt color c8, color c8}) indices.

\subsection{Declaring new specials}

Specials with indices of user-defined types can be declared by
means of
{\tt special} statement: \begin{quote} { \tt special \sl name\/\tt :(\sl
islist\tt). } \end{quote}
Here {\sl name} is the name of new symbol, and {\sl ilist} is a
comma-separated list of indices types. For example,
Gell-Mann matrices can be defined as (although color
group and its indices types are already defined):
\begin{quote} {\tt special lambda:(color c3, color c3b, color c8).
 }\end{quote}

To define Dirac's $\gamma$-matrices we can use the command
\begin{quote} {\tt special gamma:(spinor, cspinor, vector). }\end{quote}

\subsection{Arrays}

Array, i.e. the object with explicit components, can also have the
user-defined type of index. In this case generic form of such object
 is
\begin{quote} {\tt \{ \sl expr1, expr2 ... ,exprN ; itype \} }\end{quote}
where $N$ expressions {\sl expr1 ... exprN} of $N$ components are separated
 by comma,
and {\sl itype} is an optional index type.
 If {\sl itype} is
omitted LanHEP uses default group name {\tt wild} and index type
{\tt wild \sl N}, where {\sl N} is a number of components in the array.

\section{Auxiliary statements}

\subsection{Orthogonal matrices}

If some parameters appear to be the elements of the orthogonal matrix such
as quark mixing Cabbibo-Kobayashi-Maskava matrix, one should declare them
  by the statement
\begin{quote} {\tt OrthMatrix( \{\{$a_{11}$, $a_{12}$, $a_{13}$\}, 
\{$a_{21}$, $a_{22}$, $a_{23}$\}, \{$a_{31}$, $a_{32}$, $a_{33}$\}\} ).} 
\end{quote}
where $a_{ij}$ denote the parameters. Such declaration permits LanHEP
to reduce expressions which contain these parameters by taking into
account the properties of orthogonal matrices. 

Note that this statement has no relation to the arrays;
it just declares that these parameters $a_{ij}$ satisfy the correspondent
relations.
Of course, one can declare further a matrix with these parameters
as components by means of {\tt let} statement.

\subsection{Including files}

LanHEP allows the user to divide the input file into several files.
 To include the file {\sl file}, the user should
use the statement \begin{quote} {\tt read \sl file\tt . } \end{quote}
 The standard extension '{\tt .mdl}' of
the file name may be omitted in this statement.

Another way to include a file is provided by the {\tt use} statement as
\begin{quote} {\tt use  \sl file\tt . } \end{quote}
 The {\tt use} statement
reads the {\tt file} only once, next appearances of this statement with the
same argument do nothing. This function prevents multiple reading of the
 same file.
This form can be used mainly to include
some standard modules, such as declaration of Standard Model particles
to be used for writing some extensions of this model.

\subsection{Checking electric charge conservation}

LanHEP can check whether the introduced vertices satisfy electric charge
conservation law. This option is available, if the user declares some
 parameter
to denote elementary electric charge (say, {\tt ee} in QED example),
and than indicate, which particle is a photon  by the statement
\begin{quote} {\tt SetEM(\sl photon, param).}\end{quote}
So, in example of Section 1 this statement could be
\begin{quote} {\tt SetEM(A,ee).} \end{quote}

Electric charge of each particle is determined by analyzing its
 interaction 
with the photon. LanHEP
checks whether the sum of electric charges of particles in each vertex
equals zero.

\section{LaTeX output}

LanHEP generates LaTeX output instead of CompHEP model files if user 
set {\tt -tex} in the command line to start LanHEP.
Three files are produced: {\tt 'vars\sl N\/\tt .tex'}, 
{\tt 'prtcls\sl N\/\tt .tex'} and {\tt 'lgrng\sl N\/\tt .tex'}.
The first file contains names of parameter used in physical model
and their values. The second file describes introduced particles,
together with propagators derived from introduced vertices.

The last file lists introduced vertices. LanHEP uses Greece letters
$\mu, \nu, \rho$... for vector indices, letters $a,b,c$... for 
spinor ones and $p,q,r$... for color indices (and for indices of other
groups, if they were defined).

It is possible to inscribe names for particles and parameters
to use them in LaTeX output. It can be done by the statement
\begin{quote} {\tt SetTexName([ \sl ident=texname, ... \tt]).}\end{quote}
Here {\sl ident\/} is an identifier of particle or parameter, and 
{\sl texname\/} is string constant containing LaTeX command.
Note, that for introducing backslash '$\backslash$' in quoted string
 constant
one should type it twice: '$\backslash\backslash$'.

For example, if one has declared neutrino with name  {\tt n1} (and
name for antineutrino {\tt N1}) than the statement
\begin{quote} {\tt SetTexName([n1='$\backslash\backslash$nu\^{}e',
N1='$\backslash\backslash$bar\{$\backslash\backslash$nu\}\^{}e']).
}\end{quote}
makes LanHEP to use symbols $\nu^e$ and $\bar{\nu}^e$ for neutrino 
and antineutrino in LaTeX tables.

\section{Running LanHEP}

As it was mentioned above, LanHEP can read the model description 
from the input file prepared by user.
To start LanHEP write the command
\begin{quote} {\tt lhep \sl filename options} \end{quote}
where the possible {\sl options} are described in the next section.
If the {\sl filename} is omitted, LanHEP prints
it's prompt and waits for the keyboard input. In the last case,
user's input is copied into the file {\tt lhep.log } and can be inspected
in the following. To finish the work with LanHEP,
type {\tt 'quit.'} or simply press {\tt \^{}D }
(or {\tt \^{}Z} at MS DOS computers).

\subsection{Options}
Possible options, which can be used in the command line to start LanHEP
are:
\begin{itemize}
\item[ ] \underline{\tt -OutDir \sl directory\/} Set the directory where
 output files
will be placed.
\item[ ] \underline{\tt -InDir \sl directory\/} Set the default directory
 where to
search files, which included by {\tt read} and {\tt use} statements.
\item[ ] \underline{\tt -tex} LanHEP generates LaTeX files instead of 
CompHEP model tables.
\item[ ] \underline{\tt -frc} If {\tt -tex\/} option is set, forces LanHEP
 to split
4-fermion and 4-color vertices just as it is made for CompHEP files.
\item[ ] \underline{\tt -texLines \sl num\/} Set number of lines in LaTeX
tables to {\sl num}. After the specified number of lines, LanHEP continues
writing current table on the next page of LaTeX the output. Default
 value is 40.
\item[ ] \underline{\tt -texLineLength \sl num\/} Controls  width of the
 Lagrangian
table.  Default value is 35, user can vary table width by changing this
parameter.
\end{itemize}

\section{Default objects}

When LanHEP starts, it has already declared some frequently used symbols.
They are:
\begin{itemize}
\item specials {\tt gamma} and {\tt gamma5} which are Dirac's
 $\gamma$-matrices;
\item special {\tt moment} is replaced by the momentum of the corresponding
particle;
\item let-substitution {\tt deriv} is defined as {\tt -i*moment};
\item group {\tt color} with indices types {\tt c3/c3b} and {\tt c8};
\item specials {\tt lambda} and {\tt f\_SU3} which are Gell-Mann matrices
 and structure constant of SU(3) group;
\item {\tt tau1, tau2, tau3} are $\tau$-matrices $\tau_1, \tau_2, \tau_3$;
\item {\tt tau} is a vector $\vec{\tau}=(\tau_1, \tau_2, \tau_3)$;
\item {\tt taup} and {\tt taum} are matrices $\tau^\pm=(\tau^1\pm i\tau^2)/
\sqrt{2}$;
\item {\tt taupm} is a vector $(\tau^+, \tau^3, \tau^-)$;
\item {\tt eps} is the antisymmetrical tensor
                 $\varepsilon$ ($\varepsilon^{123}=1$);
\item {\tt Tau1, Tau2, Tau3} are generators of $SU(2)$ group adjoint
representation (3-dimensional analog of $\tau$-matrices)
 $T^1, T^2, T^3$
 with commutative relations $[T_i,T_j]=-i\epsilon_{ijk}T_k$: 
\[
T^1= \frac{1}{\sqrt{2}} \left(\begin{array}{rrr} 0 & -1 & 0 \\
 -1 & 0 & 1 \\ 0 & 1 & 0 \end{array} \right), \;\;\;\;
T^2= \frac{1}{\sqrt{2}}  \left(\begin{array}{rrr} 0 & i & 0 \\
 -i & 0 & -i \\ 0 & i & 0  \end{array} \right), \;\;\;\;
T^3=\left(\begin{array}{rrr} 1 & 0 & 0 \\ 0 & 0 & 0 \\ 0 & 0 & -1
  \end{array} \right);
\]
\item {\tt Taup} and {\tt Taum} corresponds to $T^\pm=(T^1\pm iT^2)/
\sqrt{2}$;
\item {\tt Taupm} is a vector $\vec{T}=(T^+, T^3, T^-)$.
\end{itemize}

\eject

\section{Standard Model}

In this chapter we give an example of the LanHEP input file
introducing the Lagrangian of the Standard Model in the t'Hooft-Feynman
gauge.

We start the description of the SM Lagrangian from the declaration of all
 parameters, 
see Fig.~\ref{fig:sm-param}. The comments which follow each parameter
explain its meaning.

\begin{figure}[h]
\framebox{ \vbox{
\flushleft \tt
model 'Standard Model'/4. \\
 parameter \\
\hspace*{1cm} EE  = 0.31333:'Elementary electric charge',\\
\hspace*{1cm} GG  = 1.117:'Strong coupling constant (Z point) (PDG-94)',\\
\hspace*{1cm} SW  
= 0.4740:'sin of the Weinberg angle (PDG-94,"on-shell")',\\
\hspace*{1cm} s12 = 0.221:'Parameter of C-K-M matrix (PDG-94)',\\
\hspace*{1cm} s23 = 0.040:'Parameter of C-K-M matrix (PDG-94)',\\
\hspace*{1cm} s13 = 0.0035:'Parameter of C-K-M matrix (PDG-94)'.\\
parameter
	CW  = sqrt(1-SW**2) : 'cos of the Weinberg angle'.\\
parameter\\
\hspace*{1cm} c12  = sqrt(1-s12**2):'parameter  of C-K-M matrix',\\
\hspace*{1cm} c23  = sqrt(1-s23**2):'parameter  of C-K-M matrix',\\
\hspace*{1cm} c13  = sqrt(1-s13**2):'parameter  of C-K-M matrix'.\\
parameter \\
\hspace*{1cm} Vud = c12*c13:'C-K-M matrix element',\\
\hspace*{1cm} Vus = s12*c13:'C-K-M matrix element',\\
\hspace*{1cm} Vub = s13:'C-K-M matrix element',\\
\hspace*{1cm} Vcd = (-s12*c23-c12*s23*s13):'C-K-M matrix element',\\
\hspace*{1cm} Vcs = (c12*c23-s12*s23*s13):'C-K-M matrix element',\\
\hspace*{1cm} Vcb = s23*c13:'C-K-M matrix element',\\
\hspace*{1cm} Vtd = (s12*s23-c12*c23*s13):'C-K-M matrix element',\\
\hspace*{1cm} Vts = (-c12*s23-s12*c23*s13):'C-K-M matrix element',\\
\hspace*{1cm} Vtb = c23*c13:'C-K-M matrix element'.\\
OrthMatrix( \{\{Vud,Vus,Vub\}, \{Vcd,Vcs,Vcb\}, \{Vtd,Vts,Vtb\}\} ).
}}
\caption{Standard Model: parameters.}
\label{fig:sm-param}
\end{figure}

Than we go to input particles of the model. These are
vector gauge bosons, fermions which are leptons and quarks of three
generations, and scalar Higgs boson (see Fig.~\ref{fig:sm-part}).

\begin{figure}[h]
\framebox{ \vbox{
\flushleft \tt
vector \\
\hspace*{1cm} A/A: (photon, gauge),\\
\hspace*{1cm} Z/Z:
 ('Z boson', mass MZ = 91.187, width wZ = 2.502, gauge),\\
\hspace*{1cm} G/G: (gluon, color c8, gauge),\\
\hspace*{1cm} 'W+'/'W-':
 ('W boson', mass MW = MZ*CW, width wW = 2.094, gauge).\\
spinor\\
\hspace*{1cm} n1:(neutrino,left),   
    e1:(electron, mass Me = 0.000511),\\
\hspace*{1cm} n2:('mu-neutrino',left),  e2:(muon, mass Mm  = 0.1057),\\
\hspace*{1cm} n3:('tau-neutrino',left),
 e3:('tau-lepton', mass Mt  = 1.777).\\
spinor\\
\hspace*{1cm} u:('u-quark',color c3),\\
\hspace*{1cm} d:('d-quark',color c3),\\
\hspace*{1cm} c:('c-quark',color c3, mass Mc  = 1.300),\\
\hspace*{1cm} s:('s-quark',color c3, mass Ms = 0.200),\\
\hspace*{1cm} t:('t-quark',color c3,
 mass Mtop = 170, width wtop = 1.442),\\
\hspace*{1cm} b:('b-quark',color c3, mass Mb =  4.300 ).\\
scalar H/H:(Higgs, mass MH = 200, width wH = 1.461).
}}
\caption{Standard Model: particles.}
\label{fig:sm-part}
\end{figure}

The next section of the input file (Fig.~\ref{fig:sm-let}) 
introduces some useful let-substitutions. First, we enter doublets of
leptons and quarks; for the quarks we supply also doublets rotated
by CKM matrix.

\begin{figure}[h]
\framebox{ \vbox{
\flushleft \tt
let l1=\{n1,e1\}, L1=\{N1,E1\}.\\
let l2=\{n2,e2\}, L2=\{N2,E2\}.\\
let l3=\{n3,e3\}, L3=\{N3,E3\}.\\
let q1=\{u,d\}, Q1=\{U,D\}, q1a=\{u,Vud*d+Vus*s+Vub*b\},
				Q1a=\{U,Vud*D+Vus*S+Vub*B\}.\\
let q2=\{c,s\}, Q2=\{C,S\}, q2a=\{c,Vcd*d+Vcs*s+Vcb*b\},
				Q2a=\{C,Vcd*D+Vcs*S+Vcb*B\}.\\
let q3=\{t,b\}, Q3=\{T,B\}, q3a=\{t,Vtd*d+Vts*s+Vtb*b\},
				Q3a=\{T,Vtd*D+Vts*S+Vtb*B\}.\\
let B1= -SW*Z+CW*A, W3=CW*Z+SW*A, W1=('W+'+'W-')/Sqrt2,
	 W2 = i*('W+'-'W-')/Sqrt2.\\
let WW=\{'W+', W3, 'W-'\}, WWc=\{'W-', W3, 'W+'\}, WW1=\{W1, W2, W3\}.\\
let gh1=('W+.c'+'W-.c')/Sqrt2, gh2= i*('W+.c'-'W-.c')/Sqrt2,
		gh3= CW*'Z.c'+SW*'A.c',\\
\hspace*{1cm} gh=\{gh1,gh2,gh3\}.\\
let Gh1 = ('W+.C'+'W-.C')/Sqrt2, Gh2=i*('W+.C'-'W-.C')/Sqrt2,
		Gh3= CW*'Z.C'+SW*'A.C',
\hspace*{1cm} Gh=\{Gh1,Gh2,Gh3\}. \\
let g=EE/SW, g1=EE/CW.\\
}}
\caption{Standard Model: useful substitutions.}
\label{fig:sm-let}
\end{figure}

In the textbooks the physical gauge bosons $W^\pm$, $Z$, $A$ are defined 
through the $SU(2)$ gauge triplet $W^a\; (a=1,2,3)$ and $U(1)$ field $B$ as
\begin{eqnarray*}
W^\pm & = & (W^1\mp iW^2)/\sqrt{2},\\
Z & = & W^3\cos \theta_W-B\sin \theta_W,\\
A & = & W^3\sin \theta_W+B\cos \theta_W,
\end{eqnarray*}
where $\theta_W$ is the Weinberg angle.
If we would like to use the fields $W^a$ and
 $B$ we have to declare them (as
let-substitutions) through the
physical fields $W^\pm$, $Z$ and $A$ by inverting these 
relations\footnote{Note, that {\tt B1} stands for the $B$ gauge field since
{\tt B} is already defined as b-antiquark.}.
At last, we
define EW charges $g=e/\sin \theta_W$ and $g_1=e/\cos \theta_W$.

Now we can enter the Lagrangian terms. First, we define the
EW gauge bosons Lagrangian (see Fig.~\ref{fig:sm-gaug}) in the form
$$ L_G^{EW} = -\frac{1}{4}F^{\mu\nu}F_{\mu\nu}
	-\frac{1}{4}W^{a\mu\nu}W^a_{\mu\nu}, $$
where
$$ F_{\mu\nu}=\partial_\mu B_\nu-\partial_\nu B_\mu \mbox{ and }
 W^a_{\mu\nu}=\partial_\mu W^a_\nu-\partial_\nu W^a_\mu-
	g\varepsilon^{abc}W^b_\mu W^c_\nu,$$
and for the gluon Lagrangian we use the form
$$ L_G^{QCD} = -\frac{1}{4}G^{a\mu\nu}G^a_{\mu\nu}, $$
where
$$
G^a_{\mu\nu}=\partial_\mu G^a_\nu-\partial_\nu G^a_\mu-
	g_s f^{abc}G^b_\mu G^c_\nu,$$
$G^a_\mu$ is the gluon field, $g_s$ is a strong interaction coupling.

\begin{figure}[h]
\framebox{ \vbox{
\flushleft \tt
let tB\^{}mu\^{}nu=deriv\^{}mu*B1\^{}nu-deriv\^{}nu*B1\^{}mu.\\
let tW\^{}mu\^{}nu\^{}a=deriv\^{}mu*WW1\^{}nu\^{}a - 
deriv\^{}nu*WW1\^{}mu\^{}a - 
g*eps\^{}a\^{}b\^{}c*WW1\^{}mu\^{}b*WW1\^{}nu\^{}c.\\
let tG\^{}mu\^{}nu\^{}a=deriv\^{}mu*G\^{}nu\^{}a -
	deriv\^{}nu*G\^{}mu\^{}a - 
 i*GG*f\^{}a\^{}b\^{}c*G\^{}mu\^{}b*G\^{}nu\^{}c.\\
lterm -tB**2/4-tW**2/4-tG**2/4.
}}
\caption{Standard Model: gauge field Lagrangians}
\label{fig:sm-gaug}
\end{figure}

The next step is the interaction of fermions with
gauge bosons (Fig.~\ref{fig:sm-ferm}).
First, we define interaction of left-handed fermions with $W^a$ and $B$
fields, taking into account CKM mixing of quarks; the corresponding term is
$$
\bar\psi\gamma^\mu\frac{1-\gamma^5}{2}\left(i\partial_\mu - 
\frac{g}{2}\tau^aW^a_\mu - g_1 Y B_\mu\right)\psi,$$ 
where $\psi$ are left-handed lepton and rotated by CKM matrix quark
doublets, $Y$ is a hypercharge.
Than we declare the right-hand fermions interaction with $B$ field;
in this case we can take quarks without CKM mixing, since
 the interaction of
quarks with $Z$-boson and photon is flavor-diagonal. Lagrangian term is
$$  \bar \psi \gamma^\mu \frac{1+\gamma^5}{2}
	(i\partial_\mu - g_1 Y B_\mu) \psi,$$
where $\psi$ are right-handed lepton and  quark singlets.

\begin{figure}[h]
\framebox{ \vbox{
\flushleft \tt
lterm  	anti(psi)*gamma*(1-g5)/2*(i*deriv-g*taupm*WW/2-Y*g1*B1)*psi
\hspace*{0.5cm} where\\ 
\hspace*{1cm} psi=l1,  Y= -1/2;\\
\hspace*{1cm} psi=l2,  Y= -1/2;\\
\hspace*{1cm} psi=l3,  Y= -1/2;\\
\hspace*{1cm} psi=q1,  Y=1/6;\\
\hspace*{1cm} psi=q2,  Y=1/6;\\
\hspace*{1cm} psi=q3,  Y=1/6.\\
lterm  	anti(psi)*gamma*(1+g5)/2*(i*deriv-Y*g1*B1)*psi\\
\hspace*{0.5cm} where \\
\hspace*{1cm} psi=e1,  Y= -1;\\
\hspace*{1cm} psi=e2, Y= -1;\\
\hspace*{1cm} psi=e3,  Y= -1;\\
\hspace*{1cm} psi=u,  Y=2/3;\\
\hspace*{1cm} psi=c,  Y=2/3;\\
\hspace*{1cm} psi=t,  Y=2/3;\\
\hspace*{1cm} psi=d,  Y=-1/3;\\
\hspace*{1cm} psi=s, Y=-1/3;\\
\hspace*{1cm} psi=b,  Y=-1/3.
}}
\caption{Standard Model: gauge interaction of fermions.}
\label{fig:sm-ferm}
\end{figure}

The next step is the introduction of the Higgs sector: interaction of Higgs
doublet expressed through physical field
$H$ and unphysical goldstone fields $Z_f, W^\pm_f$ as
$$ \Phi=\left(\begin{array}{c} -iW_f^+ \\ \frac{1}{\sqrt{2}}(v+H+iZ_f) 
\end{array} \right),$$
where vacuum expectation value is $v=2M_W/g$. Unphysical goldstone
 fields read
in LanHEP as {\tt 'W+.f', 'W-.f', 'Z.f'}. In Fig.~\ref{fig:sm-higgs} we
enter self-interaction terms and interaction of this field with gauge
 bosons as:
$$L_{Higgs} = (D_\mu\Phi)^*(D^\mu\Phi) + 2\lambda(\Phi^*\Phi-v^2/2)^2,$$
where $D_\mu=\partial_\mu+i\frac{g}{2}\tau^aW^a_\mu+i\frac{g_1}{2}B_\mu$, 
$\lambda=\frac{g^2M_H^2}{16M_W^2}$.

\begin{figure}[h]
\framebox{ \vbox{
\flushleft \tt
let phi = \{ -i*'W+.f',  (vev(2*MW/EE*SW)+H+i*'Z.f')/Sqrt2 \},\\ 
\hspace*{0.5cm} Phi = \{  i*'W-.f', (vev(2*MW/EE*SW)+H-i*'Z.f')/Sqrt2 \}.\\
lterm -2*lambda*(pp*PP+v**2/2)**2 \\
\hspace*{1cm} where lambda=(EE*MH/(MW*SW))**2/16, v=2*MW*SW/EE.\\
lterm  (DPhi)*(Dphi) where\\
\hspace*{1cm} DPhi = (deriv\^{}mu-i*g1/2*B1\^{}mu)*Phi\^{}a -
         i*g/2*taupm\^{}a\^{}b\^{}c*WWc\^{}mu\^{}c*Phi\^{}b,\\ 
\hspace*{1cm} Dphi  = = (deriv\^{}mu+i*g1/2*B1\^{}mu)*phi\^{}a +
         i*g/2*taupm\^{}a\^{}b\^{}c*WWc\^{}mu\^{}c*phi\^{}b.
}}
\caption{Standard Model: self-interaction of Higgs field and its
 interaction
with gauge bosons.}
\label{fig:sm-higgs}
\end{figure}

Yukawa terms are shown in Fig.~\ref{fig:sm-yukawa}. First, we generate
lepton masses with terms
$$ L^\ell_Y=-\frac{gM_\ell}{\sqrt{2}M_W}\left(
	\bar \psi^a\frac{1+\gamma^5}{2}\psi\Phi^a + 
        \bar\psi \frac{1-\gamma^5}{2}\psi^a\Phi^{*a}\right),$$
where $\psi^a$ is lepton doublet, $\psi$ is down lepton and $M_\ell$ is its
mass. The term is applied to the leptons of second and third generations
(electron is massless in this model).

Next we do the same thing with down quarks. Here we must take into account
the CKM mixing, so the previous formula takes the form
$$ L_Y^{Q_d}=-\frac{gM_q^{ij}}{\sqrt{2}M_W}\left(
	\bar \psi^a_i\frac{1+\gamma^5}{2}\psi_j\Phi^a + 
        \bar\psi_i \frac{1-\gamma^5}{2}\psi^a_j\Phi^{*a}\right),$$
where $M_q^{ij} = diag(M_d,M_s,M_b)^{ik}U^{jk}$, $U$ being the 
CKM matrix; $\psi^a_i$ and $\psi_i$ are quark doublets and down singlets,
indices $i,j$ numerate generations.

At last we generate masses of upper quarks with the terms
$$L^{Q_u}_Y=-\frac{gM_q}{\sqrt{2}M_W}\left(
	\bar \psi^a\frac{1+\gamma^5}{2}i\tau_2^{ab}\psi\Phi^{*b} + 
        \bar\psi \frac{1-\gamma^5}{2}\psi^ai\tau_2^{ab}\Phi^b\right)$$
for three generations of quarks with $M_q =(M_u, M_c, M_t)$. 

\begin{figure}[h]
\framebox{ \vbox{
\flushleft \tt
lterm  -M/MW/Sqrt2*g*(anti(pl)*(1+g5)/2*pr*phi + anti(pr)*(1-g5)/2*pl*Phi )
 where\\
\hspace*{1cm} M=Mm, pl=l2, pr=e2;\\
\hspace*{1cm} M=Mt, pl=l3, pr=e3.\\
lterm  -M/MW/Sqrt2*g*(anti(pl)*(1+g5)/2*pr*phi + 
anti(pr)*(1-g5)/2*pl*Phi ) where\\
\hspace*{1cm} M=Vud*0,  pl=q1a,  pr=d;  \% here and below 0 stands for Md\\
\hspace*{1cm} M=Vus*Ms, pl=q1a,  pr=s;\\
\hspace*{1cm} M=Vub*Mb, pl=q1a,  pr=b;\\
\hspace*{1cm} M=Vcd*0,  pl=q2a,  pr=d;\\
\hspace*{1cm} M=Vcs*Ms, pl=q2a,  pr=s;\\
\hspace*{1cm} M=Vcb*Mb, pl=q2a,  pr=b;\\
\hspace*{1cm} M=Vtd*0,  pl=q3a,  pr=d;\\
\hspace*{1cm} M=Vts*Ms, pl=q3a,  pr=s;\\
\hspace*{1cm} M=Vtb*Mb, pl=q3a,  pr=b.\\
lterm -M/MW/Sqrt2*g*(anti(pl)*(1+g5)/2*i*tau2*pr*Phi + 
      anti(pr)*(1-g5)/2*i*pl*tau2*phi )\\
\hspace*{2cm} where\\
\hspace*{1cm} M=0,  pl=q1a, pr=u;\\
\hspace*{1cm} M=Mc, pl=q2a, pr=c;\\
\hspace*{1cm} M=Mtop, pl=q3a,  pr=t.
}}
\caption{Standard Model: Yukawa terms.}
\label{fig:sm-yukawa}
\end{figure}

The final stage is the introduction of gauge fixing terms
 and terms with Faddev-Popov
ghosts.

Gauge fixing terms in the t'Hooft-Feynman gauge read as
$$
L_{GF} = -\frac{1}{2}(\partial_\mu A^\mu)^2 -
  \frac{1}{2}(\partial_\mu G_a^\mu)^2 - (\partial_\mu W^{+\mu}+M_W W^+_f)
(\partial_\mu W^{-\mu}+M_W W^-_f) - \frac{1}{2}(\partial_\mu Z^\mu + M_Z
 Z_f)^2,$$
the corresponding LanHEP code is shown in Fig. \ref{fig:sm-gfix}.

\begin{figure}[h]
\framebox{ \vbox{
\flushleft \tt
lterm -1/2*(deriv*A)**2. \\
lterm -1/2*(deriv*G)**2. \\
lterm -1/2*(2*(deriv*'W+'+MW*'W+.f')*(deriv*'W-'+MW*'W-.f') + \\
\hspace*{1cm}  (deriv*Z+MW/CW*'Z.f')**2).
}}
\caption{Standard Model: gauge fixing terms.}
\label{fig:sm-gfix}
\end{figure}

The interaction of ghost fields reads as (we omit here bilinear terms,
since they are not treated by CompHEP):

$$ 
L^G_{FP} = ig_s f^{abc}\bar{C}_G^a G_\mu^b \partial^\mu C_G^c $$
for gluons, and
$$
L^{EW}_{FP} = -g \varepsilon^{abc}*C^a_{W} W^b_\mu \bar{C}^c_{W} $$
for the interaction of EW ghosts with gauge bosons, and finally
\begin{eqnarray*}
L^{GS}_{FP} & = & -\frac{e M_W}{2 \sin \theta_W}
 \big((H+iZ_f)(\bar{C}{}_- C_+ 
+ \bar{C}{}_+ C_-) + \\
& & + H \bar{C}{}_Z C_Z/\cos^2 \theta_W -2iZ_f\bar{C}{}_+ C_-\big) + \\
& & +\frac{ieM_W}{2\cos \theta_W \sin \theta_W}\big(
W^+_f((1-2\sin^2\theta_W)\bar{C}_- C_Z +C_-\bar{C}{}_Z + 
2\sin\theta_W\cos\theta_W\bar{C}{}_-C_A) - \\
& & -W^-_f((1-2\sin^2\theta_W)\bar{C}_+ C_Z +C_+\bar{C}{}_Z + 
2\sin\theta_W\cos\theta_W\bar{C}{}_+C_A) \big),
\end{eqnarray*}

Here $C_G^a$ (a=1..8) are ghost fields corresponding to gluon,
$C_{W}^a$ (a=1..3) are ghost fields of $SU(2)$ gauge bosons,
 $C_B$ corresponds
to $U(1)$ gauge field and

\begin{eqnarray*}
C_\pm & = & (C^1_W\mp C^2_W)/\sqrt{2},\\
C_Z & = & C^3_W\cos \theta_W-C_B\sin \theta_W,\\
C_A & = & C^3_W\sin \theta_W+C_B\cos \theta_W.
\end{eqnarray*}

The corresponding part of LanHEP input file is
 shown in Fig. \ref{fig:sm-ghost}.
  
\begin{figure}[h]
\framebox{ \vbox{
\flushleft \tt
lterm i*GG*f*ccghost(G)*G*deriv*ghost(G).\\
lterm i*g*eps*Gh*WW*deriv*gh.\\
lterm  -1/2*(2*(deriv*'W+'+MW*'W+.f')*(deriv*'W-'+MW*'W-.f') + \\
\hspace*{1cm} (deriv*Z+MW/CW*'Z.f')**2).\\
lterm -MW*EE/2/SW*((H+i*'Z.f')*('W-.C'*'W+.c' + 'W+.C'*'W-.c')\\
\hspace*{1cm}  +H*'Z.C'*'Z.c'/CW**2-2*i*'Z.f'*'W+.C'*'W-.c').
lterm i*EE*MW/2/CW/SW*('W+.f'*('W-.C'*'Z.c'*(1-2*SW**2)+'W-.c'*'Z.C'
\hspace*{2cm}      +2*CW*SW*'W-.C'*'A.c') - \\
\hspace*{1cm}   'W-.f'*('W+.C'*'Z.c'*(1-2*SW**2)+'W+.c'*'Z.C'\\
\hspace*{2cm}     +2*CW*SW*'W+.C'*'A.c')).
}}
\caption{Standard Model: terms with Faddev-Popov ghost fields.}
\label{fig:sm-ghost}
\end{figure}

To complete the description of Standard Model, we prescribe also LaTeX
names for particles and parameters, see Fig. \ref{fig:sm-tex}.

\begin{figure}[h]
\framebox{ \vbox{
\flushleft \tt
SetTexName([u=u, U='$\backslash\backslash$bar\{u\}', d=d, 
D='$\backslash\backslash$bar\{d\}']).\\
SetTexName([s=s, S='$\backslash\backslash$bar\{s\}', b=b, 
B='$\backslash\backslash$bar{b}']).\\
SetTexName([c=c, C='$\backslash\backslash$bar\{c\}', t=t, 
T='$\backslash\backslash$bar\{t\}']).\\
SetTexName([e1=e, E1='$\backslash\backslash$bar\{e\}', 
n1='$\backslash\backslash$nu\^{}e', 
N1='$\backslash\backslash$bar\{$\backslash\backslash$nu\}\^{}e']).\\
SetTexName([e2='$\backslash\backslash$mu', 
E2='$\backslash\backslash$bar\{$\backslash\backslash$mu\}', 
n2='$\backslash\backslash$nu\^{}$\backslash\backslash$mu', 
N2='$\backslash
\backslash$bar\{$\backslash
\backslash$nu\}\^{}$\backslash\backslash$mu']).\\
SetTexName([e3='$\backslash\backslash$tau', 
E3='$\backslash\backslash$bar\{$\backslash\backslash$tau\}', 
n3='$\backslash\backslash$nu\^{}$\backslash\backslash$tau', 
N3='$\backslash\backslash$bar\{$\backslash\backslash$nu\}\^{}$\backslash
\backslash$tau']).\\
SetTexName([EE=e, GG='g\_s', SW='s\_w', CW='c\_w', MZ='M\_Z']).\\
SetTexName([Me='M\_e', Mm='M\_$\backslash\backslash$mu',
Mt='M\_$\backslash\backslash$tau', Ms='M\_s',
                Mc='M\_c', Mb=M\_b, Mtop=M\_t]).
}}
\caption{Standard Model: LaTeX names for particles and parameters.}
\label{fig:sm-tex}
\end{figure}


\eject

\appendix
{\huge \bf Appendix}

\section{Processing 4-color vertices \label{imp}}

CompHEP Lagrangian tables don't describe  explicitly color
structure of a vertex. If color particles present in the vertex,
the following implicit convolutions are assumed (supposing $p,q,r$
are color indices of particles in the vertex):
\begin{itemize}
\item $\delta_{pq}$ for two color particles $A^1_p$, $A^2_q$;
\item $\lambda_{pq}^r$ for three particles, which are color triplet,
antitriplet and octet;
\item $f^{pqr}$ for three color octets.
\end{itemize}
Other color structures are forbidden in CompHEP.

So, to introduce the 4-gluon vertex 
$f^{pqr}G_\mu^qG_\nu^rf^{pst}G_\mu^sG_\nu^t$ one should
split this 4-legs vertex into 3-legs vertices 
$f^{pqr}G_\mu^qG_\nu^rX_{\mu\nu}^p$:

{\scriptsize
\linethickness{0.5pt}
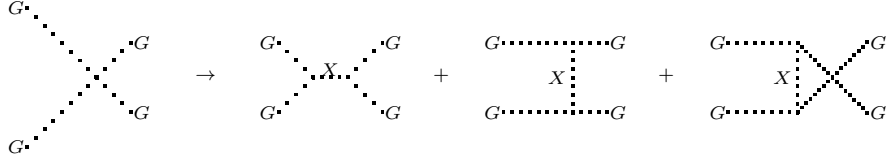
\begin{center}
\begin{picture}(62,77)(0,0)
\put(9.9,70.2){\makebox(0,0)[r]{$G$}}
\multiput(10.4,70.2)(3.3,-3.3){9}{\rule[-0.5pt]{1.0pt}{1.0pt}}
\put(9.9,17.8){\makebox(0,0)[r]{$G$}}
\multiput(10.4,17.8)(3.3,3.3){9}{\rule[-0.5pt]{1.0pt}{1.0pt}}
\put(51.1,57.1){\makebox(0,0)[l]{$G$}}
\multiput(36.5,44.0)(3.3,3.3){5}{\rule[-0.5pt]{1.0pt}{1.0pt}}
\put(51.1,30.9){\makebox(0,0)[l]{$G$}}
\multiput(36.5,44.0)(3.3,-3.3){5}{\rule[-0.5pt]{1.0pt}{1.0pt}}
\end{picture} \ 
\begin{picture}(25,77)(0,0)
\put(7,43){$\rightarrow$}
\end{picture}
\begin{picture}(62,77)(0,0)
\put(9.9,57.1){\makebox(0,0)[r]{$G$}}
\multiput(10.4,57.1)(3.3,-3.3){5}{\rule[-0.5pt]{1.0pt}{1.0pt}}
\put(9.9,30.9){\makebox(0,0)[r]{$G$}}
\multiput(10.4,30.9)(3.3,3.3){5}{\rule[-0.5pt]{1.0pt}{1.0pt}}
\put(30.2,47.1){\makebox(0,0){$X$}}
\multiput(23.5,44.0)(3.3,0.0){5}{\rule[-0.5pt]{1.0pt}{1.0pt}}
\put(51.1,57.1){\makebox(0,0)[l]{$G$}}
\multiput(36.5,44.0)(3.3,3.3){5}{\rule[-0.5pt]{1.0pt}{1.0pt}}
\put(51.1,30.9){\makebox(0,0)[l]{$G$}}
\multiput(36.5,44.0)(3.3,-3.3){5}{\rule[-0.5pt]{1.0pt}{1.0pt}}
\end{picture} \
\begin{picture}(15,77)(0,0)
\put(2,43){$+$}
\end{picture} 
\begin{picture}(62,77)(0,0)
\put(9.9,57.1){\makebox(0,0)[r]{$G$}}
\multiput(10.4,57.1)(3.3,0.0){9}{\rule[-0.5pt]{1.0pt}{1.0pt}}
\put(51.1,57.1){\makebox(0,0)[l]{$G$}}
\multiput(36.5,57.1)(3.3,0.0){5}{\rule[-0.5pt]{1.0pt}{1.0pt}}
\put(34.9,44.0){\makebox(0,0)[r]{$X$}}
\multiput(36.5,57.1)(0.0,-3.3){9}{\rule[-0.5pt]{1.0pt}{1.0pt}}
\put(9.9,30.9){\makebox(0,0)[r]{$G$}}
\multiput(10.4,30.9)(3.3,0.0){9}{\rule[-0.5pt]{1.0pt}{1.0pt}}
\put(51.1,30.9){\makebox(0,0)[l]{$G$}}
\multiput(36.5,30.9)(3.3,0.0){5}{\rule[-0.5pt]{1.0pt}{1.0pt}}
\end{picture} \ 
\begin{picture}(15,77)(0,0)
\put(2,43){$+$}
\end{picture} 
\begin{picture}(62,77)(0,0)
\put(9.9,57.1){\makebox(0,0)[r]{$G$}}
\multiput(10.4,57.1)(3.3,0.0){9}{\rule[-0.5pt]{1.0pt}{1.0pt}}
\put(64.5,57.1){\makebox(0,0)[l]{$G$}}
\multiput(36.5,57.1)(1.65,-1.65){17}{\rule[-0.5pt]{1.0pt}{1.0pt}}
\put(34.9,44.0){\makebox(0,0)[r]{$X$}}
\multiput(36.5,57.1)(0.0,-3.3){9}{\rule[-0.5pt]{1.0pt}{1.0pt}}
\put(9.9,30.9){\makebox(0,0)[r]{$G$}}
\multiput(10.4,30.9)(3.3,0.0){9}{\rule[-0.5pt]{1.0pt}{1.0pt}}
\put(64.5,30.9){\makebox(0,0)[l]{$G$}}
\multiput(36.5,30.9)(1.65,1.65){17}{\rule[-0.5pt]{1.0pt}{1.0pt}}
\end{picture} \ 
\end{center}
}

Here the field $X_{\mu\nu}^p$ is Lorentz tensor and color octet,
and this field also has constant propagator. If gluon name in CompHEP is 
{\tt 'G'}, one can use name {\tt 'G.t'} for this tensor particle;
its indices denoted as {\tt 'm\_'} and {\tt 'M\_'} ({\tt '\_'} is
the number of the particle in table item).


\section{Feynman rules for the Standard Model}

Here we present the tables generated by LanHEP in LaTeX format for the
Standard Model (see Section 12).

\vskip 5mm

\begin{table}[h] 
\caption{Standard model: particles.}
\begin{center}
\begin{tabular}{|cc|l|c|c|c|l|l|} \hline
P & aP & Name & Spin  & EM charge & Color &
 2-legs vertex & Comment \\ \hline
$A_{\mu }$&$A_{\nu }$&photon        &1   
        & $\phantom{-}0$ &1    &$-p_1^\rho p_1^\rho g^{\mu \nu} $&gauge\\
$Z_{\mu }$&$Z_{\nu }$&Z boson       &1     
      & $\phantom{-}0$ &1   
 &$-\frac{1}{ c_w^2 }\left(c_w^2 p_1^\rho
 p_1^\rho g^{\mu \nu} - MW^2 \cdot g^{\mu \nu} \right)$&gauge\\
$G_{\mu p}$&$G_{\nu q}$&gluon         &1  
         & $\phantom{-}0$ &8    &$-p_1^\rho
 p_1^\rho g^{\mu \nu} \delta_{p q} $&gauge\\
$W^+{}_{\mu }$&$W^-{}_{\nu }$&W boson       &1
           &$\phantom{-}1$ &1    &$-g^{\mu \nu}
 \left(p_1^\rho p_1^\rho - MW^2 \right)$&gauge\\
$\nu^e{}_{a}$&$\bar{\nu}^e{}_{b}$&neutrino     
 &$1/2$       & $\phantom{-}0$ &1    &$-\frac{1}{2}p_1^\mu
 \left(\gamma_{a b}^\mu -\gamma_{a c}^\mu \gamma_{c b}^5 \right)$&left\\
$e{}_{a}$ &$\bar{e}{}_{b}$&electron      &$1/2$  
     &$-1$ &1    &$-p_1^\mu \gamma_{a b}^\mu $&   \\
$\nu^\mu{}_{a}$&$\bar{\nu}^\mu{}_{b}$&mu-neutrino 
  &$1/2$       & $\phantom{-}0$ &1    &$-\frac{1}{2}p_1^\mu
 \left(\gamma_{a b}^\mu -\gamma_{a c}^\mu \gamma_{c b}^5 \right)$&left\\
$\mu{}_{a}$&$\bar{\mu}{}_{b}$&muon          &$1/2$   
    &$-1$ &1    &$-\left(p_1^\mu \gamma_{a b}^\mu + M_\mu\cdot
 \delta_{a b} \right)$&   \\
$\nu^\tau{}_{a}$&$\bar{\nu}^\tau{}_{b}$&tau-neutrino 
 &$1/2$       & $\phantom{-}0$ &1    &$-\frac{1}{2}p_1^\mu
 \left(\gamma_{a b}^\mu -\gamma_{a c}^\mu \gamma_{c b}^5 \right)$&left\\
$\tau{}_{a}$&$\bar{\tau}{}_{b}$&tau-lepton    &$1/2$ 
      &$-1$ &1    &$-\left(p_1^\mu \gamma_{a b}^\mu
 + M_\tau\cdot \delta_{a b} \right)$&   \\
$u{}_{ap}$&$\bar{u}{}_{bq}$&u-quark       &$1/2$ 
      &$\phantom{-}\frac{2}{3}$ &3    &$-p_1^\mu 
\delta_{p q} \gamma_{a b}^\mu $&   \\
$d{}_{ap}$&$\bar{d}{}_{bq}$&d-quark   &$1/2$  
     &$-\frac{1}{3}$ &3    &$-\frac{1}{2}p_1^\mu
 \delta_{p q} 2\gamma_{a b}^\mu $&   \\
$c{}_{ap}$&$\bar{c}{}_{bq}$&c-quark  &$1/2$  
     &$\phantom{-}\frac{2}{3}$ &3    &$-\delta_{p q}
 \left(p_1^\mu \gamma_{a b}^\mu + M_c\cdot \delta_{a b} \right)$&   \\
$s{}_{ap}$&$\bar{s}{}_{bq}$&s-quark  &$1/2$  
     &$-\frac{1}{3}$ &3    &$-\frac{1}{2}\delta_{p q}
 \left(2p_1^\mu \gamma_{a b}^\mu +2 M_s\cdot \delta_{a b} \right)$&   \\
$t{}_{ap}$&$\bar{t}{}_{bq}$&t-quark   &$1/2$ 
      &$\phantom{-}\frac{2}{3}$ &3    &$-\delta_{p q}
 \left(p_1^\mu \gamma_{a b}^\mu + M_t\cdot \delta_{a b} \right)$&   \\
$b{}_{ap}$&$\bar{b}{}_{bq}$&b-quark   &$1/2$       
&$-\frac{1}{3}$ &3    &$-\frac{1}{2}\delta_{p q}
 \left(2p_1^\mu \gamma_{a b}^\mu +2 M_b\cdot \delta_{a b} \right)$&   \\
$H_{}$    &$H_{}$    &Higgs         &0       
    & $\phantom{-}0$ &1    &$-\left( MH^2 -p_1^\mu p_1^\mu
 \right)$&   \\ \hline
\end{tabular}
\end{center}
\end{table}
\eject

\begin{table}[h]
\caption{Standard Model: parameters.}
\begin{center}
\begin{tabular}{|l|l|l|} \hline
Parameter & Value & Comment \\ \hline
EE    &0.31333             &Electromagnetic coupling constant (1/128) \\
GG    &1.117               &Strong coupling constant (Z point)  (PDG-94) \\
SW    &0.474            
   &sin of the Weinberg angle (PDG-94,"on-shell") \\
s12   &0.221               &Parameter of C-K-M matrix (PDG-94) \\
s23   &0.04                &Parameter of C-K-M matrix (PDG-94) \\
s13   &0.0035              &Parameter of C-K-M matrix (PDG-94) \\
CW    &Sqrt(1-SW**2)       &cos of the Weinberg angle \\
c12   &Sqrt(1-s12**2)      &parameter  of C-K-M matrix \\
c23   &Sqrt(1-s23**2)      &parameter  of C-K-M matrix \\
c13   &Sqrt(1-s13**2)      &parameter  of C-K-M matrix \\
Vud   &c12*c13             &C-K-M matrix element \\
Vus   &s12*c13             &C-K-M matrix element \\
Vub   &s13                 &C-K-M matrix element \\
Vcd   &-s12*c23-c12*s23*s13&C-K-M matrix element \\
Vcs   &c12*c23-s12*s23*s13 &C-K-M matrix element \\
Vcb   &s23*c13             &C-K-M matrix element \\
Vtd   &s12*s23-c12*c23*s13 &C-K-M matrix element \\
Vts   &-c12*s23-s12*c23*s13&C-K-M matrix element \\
Vtb   &c23*c13             &C-K-M matrix element \\
MZ    &91.187              &mass of Z boson \\
wZ    &2.502               &width of Z boson \\
MW    &MZ*CW               &mass of W boson \\
wW    &2.094               &width of W boson \\
Me    &0.000511            &mass of electron \\
Mm    &0.1057              &mass of muon \\
Mt    &1.777               &mass of tau-lepton \\
Mc    &1.3                 &mass of c-quark \\
Ms    &0.2                 &mass of s-quark \\
Mtop  &170                 &mass of t-quark \\
wtop  &1.442               &width of t-quark \\
Mb    &4.3                 &mass of b-quark \\
MH    &200                 &mass of Higgs \\
wH    &1.461               &width of Higgs \\ \hline
\end{tabular}
\end{center}
\end{table}

\begin{center}
\begin{tabular}{|llll|l|} \hline
\multicolumn{4}{|c|}{Fields in the vertex} & Variational
 derivative of Lagrangian by fields \\ \hline
${G}_{\mu p }$ & ${G}_{\nu q }$ & ${G}_{\rho r }$ &
  & $ g_sf_{p q r} \big(p_3^\nu g^{\mu \rho} -p_2^\rho
 g^{\mu \nu} -p_3^\mu g^{\nu \rho} +p_1^\rho g^{\mu
 \nu} +p_2^\mu g^{\nu \rho} -p_1^\nu g^{\mu \rho}
 \big)$\\[2mm]
${G}_{\mu p }$ & ${G}_{\nu q }$ & ${G}_{\rho r }$ &
 ${G}_{\sigma s }$ & $ g_s^2 \big(g^{\mu \rho} g^{\nu
 \sigma} f_{p q t} f_{r s t} -g^{\mu \sigma} g^{\nu
 \rho} f_{p q t} f_{r s t} +g^{\mu \nu} g^{\rho
 \sigma} f_{p r t} f_{q s t} $ \\[2mm]
 & & & & $+g^{\mu \nu} g^{\rho \sigma} f_{p s t} f_{q r t}
 -g^{\mu \sigma} g^{\nu \rho} f_{p r t} f_{q s t} -g^{\mu
 \rho} g^{\nu \sigma} f_{p s t} f_{q r t} \big)$\\[2mm]
$W^+{}_{\mu }$ & $W^-{}_{\nu }$ & ${Z}_{\rho }$ &
  & $-\frac{ c_w e}{ s_w}\big(p_1^\nu g^{\mu \rho} -p_1^\rho
 g^{\mu \nu} -p_2^\mu g^{\nu \rho} +p_2^\rho g^{\mu \nu}
 +p_3^\mu g^{\nu \rho} -p_3^\nu g^{\mu \rho} \big)$\\[2mm]
${A}_{\mu }$ & $W^+{}_{\nu }$ & $W^-{}_{\rho }$ &
  & $- e\big(p_2^\rho g^{\mu \nu} -p_2^\mu g^{\nu \rho}
 -p_3^\nu g^{\mu \rho} +p_3^\mu g^{\nu \rho} +p_1^\nu
 g^{\mu \rho} -p_1^\rho g^{\mu \nu} \big)$\\[2mm]
$W^+{}_{\mu }$ & $W^-{}_{\nu }$ & ${Z}_{\rho }$ &
 ${Z}_{\sigma }$ & $-\frac{ c_w^2  e^2 }{ s_w^2 }\big(2g^{\mu
 \nu} g^{\rho \sigma} -g^{\mu \rho} g^{\nu \sigma} -g^{\mu
 \sigma} g^{\nu \rho} \big)$\\[2mm]
${A}_{\mu }$ & $W^+{}_{\nu }$ & $W^-{}_{\rho }$ &
 ${Z}_{\sigma }$ & $-\frac{ c_w e^2 }{ s_w}\big(2g^{\mu \sigma}
 g^{\nu \rho} -g^{\mu \nu} g^{\rho \sigma} -g^{\mu \rho} g^{\nu
 \sigma} \big)$\\[2mm]
${A}_{\mu }$ & ${A}_{\nu }$ & $W^+{}_{\rho }$ & $W^-{}_{\sigma }$
 & $- e^2 \big(2g^{\mu \nu} g^{\rho \sigma} -g^{\mu \rho}
 g^{\nu \sigma} -g^{\mu \sigma} g^{\nu \rho} \big)$\\[2mm]
$W^+{}_{\mu }$ & $W^+{}_{\nu }$ & $W^-{}_{\rho }$ &
 $W^-{}_{\sigma }$ & $\frac{ e^2 }{ s_w^2 }\big(2g^{\mu \nu}
 g^{\rho \sigma} -g^{\mu \sigma} g^{\nu \rho} -g^{\mu \rho}
 g^{\nu \sigma} \big)$\\[2mm]
$\bar{\nu}^e{}_{a }$ & $\nu^e{}_{b }$ & ${Z}_{\mu }$ &
  & $-\frac{1}{4}\frac{ e}{ c_w s_w}\big(\gamma_{a b}^\mu
 -\gamma_{a c}^\mu \gamma_{c b}^5 \big)$\\[2mm]
$\bar{\nu}^e{}_{a }$ & $e{}_{b }$ & $W^+{}_{\mu }$ &
  & $-\frac{1}{4}\frac{ e\sqrt{2}}{ s_w}\big(\gamma_{a
 b}^\mu -\gamma_{a c}^\mu \gamma_{c b}^5 \big)$\\[2mm]
$\bar{e}{}_{a }$ & $\nu^e{}_{b }$ & $W^-{}_{\mu }$ &
  & $-\frac{1}{4}\frac{ e\sqrt{2}}{ s_w}\big(\gamma_{a
 b}^\mu -\gamma_{a c}^\mu \gamma_{c b}^5 \big)$\\[2mm]
$\bar{e}{}_{a }$ & $e{}_{b }$ & ${Z}_{\mu }$ & 
 & $\frac{1}{4}\frac{ e}{ c_w s_w}\big(-4 s_w^2
 \gamma_{a b}^\mu +\gamma_{a b}^\mu -\gamma_{a c}^\mu \gamma_{c
 b}^5 \big)$\\[2mm]
$\bar{e}{}_{a }$ & $e{}_{b }$ & ${A}_{\mu }$ &
  & $ e\gamma_{a b}^\mu $\\[2mm]
$\bar{\nu}^\mu{}_{a }$ & $\nu^\mu{}_{b }$ & ${Z}_{\mu }$ &
  & $-\frac{1}{4}\frac{ e}{ c_w s_w}\big(\gamma_{a b}^\mu
 -\gamma_{a c}^\mu \gamma_{c b}^5 \big)$\\[2mm]
$\bar{\nu}^\mu{}_{a }$ & $\mu{}_{b }$ & $W^+{}_{\mu }$ &
  & $-\frac{1}{4}\frac{ e\sqrt{2}}{ s_w}\big(\gamma_{a b}^\mu
 -\gamma_{a c}^\mu \gamma_{c b}^5 \big)$\\[2mm]
$\bar{\mu}{}_{a }$ & $\nu^\mu{}_{b }$ & $W^-{}_{\mu }$ &
  & $-\frac{1}{4}\frac{ e\sqrt{2}}{ s_w}\big(\gamma_{a b}^\mu
 -\gamma_{a c}^\mu \gamma_{c b}^5 \big)$\\[2mm]
$\bar{\mu}{}_{a }$ & $\mu{}_{b }$ & ${Z}_{\mu }$ & 
 & $\frac{1}{4}\frac{ e}{ c_w s_w}\big(-4 s_w^2 \gamma_{a b}^\mu
 +\gamma_{a b}^\mu -\gamma_{a c}^\mu \gamma_{c b}^5 \big)$\\[2mm]
$\bar{\mu}{}_{a }$ & $\mu{}_{b }$ & ${A}_{\mu }$ &
  & $ e\gamma_{a b}^\mu $\\[2mm]
$\bar{\nu}^\tau{}_{a }$ & $\nu^\tau{}_{b }$ & ${Z}_{\mu }$ &
  & $-\frac{1}{4}\frac{ e}{ c_w s_w}\big(\gamma_{a b}^\mu
 -\gamma_{a c}^\mu \gamma_{c b}^5 \big)$\\[2mm]
$\bar{\nu}^\tau{}_{a }$ & $\tau{}_{b }$ & $W^+{}_{\mu
 }$ &  & $-\frac{1}{4}\frac{ e\sqrt{2}}{ s_w}\big(\gamma_{a b}^\mu
 -\gamma_{a c}^\mu \gamma_{c b}^5 \big)$\\[2mm]
$\bar{\tau}{}_{a }$ & $\nu^\tau{}_{b }$ & $W^-{}_{\mu
 }$ &  & $-\frac{1}{4}\frac{ e\sqrt{2}}{ s_w}\big(\gamma_{a b}^\mu
 -\gamma_{a c}^\mu \gamma_{c b}^5 \big)$\\[2mm]
$\bar{\tau}{}_{a }$ & $\tau{}_{b }$ & ${Z}_{\mu }$ &
  & $\frac{1}{4}\frac{ e}{ c_w s_w}\big(-4 s_w^2 \gamma_{a
 b}^\mu +\gamma_{a b}^\mu -\gamma_{a c}^\mu \gamma_{c b}^5 \big)$\\[2mm]
$\bar{\tau}{}_{a }$ & $\tau{}_{b }$ & ${A}_{\mu }$ &
  & $ e\gamma_{a b}^\mu $\\[2mm]
$\bar{u}{}_{a p }$ & $u{}_{b q }$ & ${Z}_{\mu }$ & 
 & $-\frac{1}{12}\frac{ e}{ c_w s_w}\delta_{p q} \big(-8 s_w^2
 \gamma_{a b}^\mu +3\gamma_{a b}^\mu -3\gamma_{a c}^\mu
 \gamma_{c b}^5 \big)$\\[2mm]
$\bar{u}{}_{a p }$ & $u{}_{b q }$ & ${A}_{\mu }$ & 
 & $-\frac{2}{3} e\delta_{p q} \gamma_{a b}^\mu $\\[2mm]
$\bar{u}{}_{a p }$ & $d{}_{b q }$ & $W^+{}_{\mu }$ & 
 & $-\frac{1}{4}\frac{ e\sqrt{2} Vud}{ s_w}\delta_{p q}
 \big(\gamma_{a b}^\mu -\gamma_{a c}^\mu \gamma_{c b}^5 \big)$\\[2mm]
$\bar{u}{}_{a p }$ & $s{}_{b q }$ & $W^+{}_{\mu }$ &
  & $-\frac{1}{4}\frac{ e\sqrt{2} Vus}{ s_w}\delta_{p
 q} \big(\gamma_{a b}^\mu -\gamma_{a c}^\mu \gamma_{c
 b}^5 \big)$\\[2mm]
$\bar{u}{}_{a p }$ & $b{}_{b q }$ & $W^+{}_{\mu }$ & 
 & $-\frac{1}{4}\frac{ e\sqrt{2} Vub}{ s_w}\delta_{p
 q} \big(\gamma_{a b}^\mu -\gamma_{a c}^\mu \gamma_{c b}^5 \big)$\\[2mm]
$\bar{d}{}_{a p }$ & $u{}_{b q }$ & $W^-{}_{\mu }$ &
  & $-\frac{1}{4}\frac{ e\sqrt{2} Vud}{ s_w}\delta_{p
 q} \big(\gamma_{a b}^\mu -\gamma_{a c}^\mu \gamma_{c b}^5 \big)$\\[2mm]
$\bar{d}{}_{a p }$ & $d{}_{b q }$ & ${Z}_{\mu }$ &
  & $\frac{1}{12}\frac{ e}{ c_w s_w}\delta_{p q}
 \big(-4 s_w^2 \gamma_{a b}^\mu +3\gamma_{a b}^\mu
 -3\gamma_{a c}^\mu \gamma_{c b}^5 \big)$\\[2mm]
$\bar{d}{}_{a p }$ & $d{}_{b q }$ & ${A}_{\mu }$ &
  & $\frac{1}{6} e\delta_{p q} 2\gamma_{a b}^\mu $\\[2mm]
$\bar{s}{}_{a p }$ & $u{}_{b q }$ & $W^-{}_{\mu }$ &
  & $-\frac{1}{4}\frac{ e\sqrt{2} Vus}{ s_w}\delta_{p
 q} \big(\gamma_{a b}^\mu -\gamma_{a c}^\mu \gamma_{c b}^5 \big)$\\[2mm]
$\bar{s}{}_{a p }$ & $s{}_{b q }$ & ${Z}_{\mu }$ &
  & $\frac{1}{12}\frac{ e}{ c_w s_w}\delta_{p q}
 \big(-4 s_w^2 \gamma_{a b}^\mu +3\gamma_{a b}^\mu
 -3\gamma_{a c}^\mu \gamma_{c b}^5 \big)$\\[2mm]
$\bar{s}{}_{a p }$ & $s{}_{b q }$ & ${A}_{\mu }$ &
  & $\frac{1}{6} e\delta_{p q} 2\gamma_{a b}^\mu $\\[2mm]
$\bar{b}{}_{a p }$ & $u{}_{b q }$ & $W^-{}_{\mu }$ &
  & $-\frac{1}{4}\frac{ e\sqrt{2} Vub}{ s_w}\delta_{p
 q} \big(\gamma_{a b}^\mu -\gamma_{a c}^\mu \gamma_{c b}^5 \big)$\\[2mm]
$\bar{b}{}_{a p }$ & $b{}_{b q }$ & ${Z}_{\mu }$ & 
 & $\frac{1}{12}\frac{ e}{ c_w s_w}\delta_{p q} \big(-4
 s_w^2 \gamma_{a b}^\mu +3\gamma_{a b}^\mu -3\gamma_{a
 c}^\mu \gamma_{c b}^5 \big)$\\ \hline
\end{tabular}

\begin{tabular}{|llll|l|} \hline
\multicolumn{4}{|c|}{Fields in the vertex} & Variational derivative
 of Lagrangian terms by fields \\ \hline
$\bar{b}{}_{a p }$ & $b{}_{b q }$ & ${A}_{\mu }$ &
  & $\frac{1}{6} e\delta_{p q} 2\gamma_{a b}^\mu $\\[2mm]
$\bar{c}{}_{a p }$ & $c{}_{b q }$ & ${Z}_{\mu }$ &
  & $-\frac{1}{12}\frac{ e}{ c_w s_w}\delta_{p q}
 \big(-8 s_w^2 \gamma_{a b}^\mu +3\gamma_{a b}^\mu
 -3\gamma_{a c}^\mu \gamma_{c b}^5 \big)$\\[2mm]
$\bar{c}{}_{a p }$ & $c{}_{b q }$ & ${A}_{\mu }$ &
  & $-\frac{2}{3} e\delta_{p q} \gamma_{a b}^\mu $\\[2mm]
$\bar{c}{}_{a p }$ & $d{}_{b q }$ & $W^+{}_{\mu }$ & 
 & $-\frac{1}{4}\frac{ e\sqrt{2} Vcd}{ s_w}\delta_{p q}
 \big(\gamma_{a b}^\mu -\gamma_{a c}^\mu \gamma_{c b}^5 \big)$\\[2mm]
$\bar{c}{}_{a p }$ & $s{}_{b q }$ & $W^+{}_{\mu }$ &
  & $-\frac{1}{4}\frac{ e\sqrt{2} Vcs}{ s_w}\delta_{p
 q} \big(\gamma_{a b}^\mu -\gamma_{a c}^\mu \gamma_{c b}^5 \big)$\\[2mm]
$\bar{c}{}_{a p }$ & $b{}_{b q }$ & $W^+{}_{\mu }$ & 
 & $-\frac{1}{4}\frac{ e\sqrt{2} Vcb}{ s_w}\delta_{p
 q} \big(\gamma_{a b}^\mu -\gamma_{a c}^\mu \gamma_{c b}^5 \big)$\\[2mm]
$\bar{d}{}_{a p }$ & $c{}_{b q }$ & $W^-{}_{\mu }$ & 
 & $-\frac{1}{4}\frac{ e\sqrt{2} Vcd}{ s_w}\delta_{p
 q} \big(\gamma_{a b}^\mu -\gamma_{a c}^\mu \gamma_{c b}^5 \big)$\\[2mm]
$\bar{s}{}_{a p }$ & $c{}_{b q }$ & $W^-{}_{\mu }$ & 
 & $-\frac{1}{4}\frac{ e\sqrt{2} Vcs}{ s_w}\delta_{p
 q} \big(\gamma_{a b}^\mu -\gamma_{a c}^\mu \gamma_{c b}^5 \big)$\\[2mm]
$\bar{b}{}_{a p }$ & $c{}_{b q }$ & $W^-{}_{\mu }$ &
  & $-\frac{1}{4}\frac{ e\sqrt{2} Vcb}{ s_w}\delta_{p
 q} \big(\gamma_{a b}^\mu -\gamma_{a c}^\mu \gamma_{c b}^5 \big)$\\[2mm]
$\bar{t}{}_{a p }$ & $t{}_{b q }$ & ${Z}_{\mu }$ & 
 & $-\frac{1}{12}\frac{ e}{ c_w s_w}\delta_{p q}
 \big(-8 s_w^2 \gamma_{a b}^\mu +3\gamma_{a b}^\mu
 -3\gamma_{a c}^\mu \gamma_{c b}^5 \big)$\\[2mm]
$\bar{t}{}_{a p }$ & $t{}_{b q }$ & ${A}_{\mu }$ &
  & $-\frac{2}{3} e\delta_{p q} \gamma_{a b}^\mu $\\[2mm]
$\bar{t}{}_{a p }$ & $d{}_{b q }$ & $W^+{}_{\mu }$ & 
 & $-\frac{1}{4}\frac{ e\sqrt{2} Vtd}{ s_w}\delta_{p
 q} \big(\gamma_{a b}^\mu -\gamma_{a c}^\mu \gamma_{c b}^5 \big)$\\[2mm]
$\bar{t}{}_{a p }$ & $s{}_{b q }$ & $W^+{}_{\mu }$ & 
 & $-\frac{1}{4}\frac{ e\sqrt{2} Vts}{ s_w}\delta_{p 
q} \big(\gamma_{a b}^\mu -\gamma_{a c}^\mu \gamma_{c b}^5 \big)$\\[2mm]
$\bar{t}{}_{a p }$ & $b{}_{b q }$ & $W^+{}_{\mu }$ & 
 & $-\frac{1}{4}\frac{ e\sqrt{2} Vtb}{ s_w}\delta_{p
 q} \big(\gamma_{a b}^\mu -\gamma_{a c}^\mu \gamma_{c b}^5 \big)$\\[2mm]
$\bar{d}{}_{a p }$ & $t{}_{b q }$ & $W^-{}_{\mu }$ & 
 & $-\frac{1}{4}\frac{ e\sqrt{2} Vtd}{ s_w}\delta_{p
 q} \big(\gamma_{a b}^\mu -\gamma_{a c}^\mu \gamma_{c
 b}^5 \big)$\\[2mm]
$\bar{s}{}_{a p }$ & $t{}_{b q }$ & $W^-{}_{\mu }$ & 
 & $-\frac{1}{4}\frac{ e\sqrt{2} Vts}{ s_w}\delta_{p
 q} \big(\gamma_{a b}^\mu -\gamma_{a c}^\mu \gamma_{c b}^5 \big)$\\[2mm]
$\bar{b}{}_{a p }$ & $t{}_{b q }$ & $W^-{}_{\mu }$ & 
 & $-\frac{1}{4}\frac{ e\sqrt{2} Vtb}{ s_w}\delta_{p 
q} \big(\gamma_{a b}^\mu -\gamma_{a c}^\mu \gamma_{c b}^5 \big)$\\[2mm]
$\bar{u}{}_{a p }$ & $u{}_{b q }$ & ${G}_{\mu r }$ & 
 & $ g_s\gamma_{a b}^\mu \lambda_{p q}^r $\\[2mm]
$\bar{d}{}_{a p }$ & $d{}_{b q }$ & ${G}_{\mu r }$ &
  & $ g_s\gamma_{a b}^\mu \lambda_{p q}^r $\\[2mm]
$\bar{c}{}_{a p }$ & $c{}_{b q }$ & ${G}_{\mu r }$ & 
 & $ g_s\gamma_{a b}^\mu \lambda_{p q}^r $\\[2mm]
$\bar{s}{}_{a p }$ & $s{}_{b q }$ & ${G}_{\mu r }$ & 
 & $ g_s\gamma_{a b}^\mu \lambda_{p q}^r $\\[2mm]
$\bar{t}{}_{a p }$ & $t{}_{b q }$ & ${G}_{\mu r }$ & 
 & $ g_s\gamma_{a b}^\mu \lambda_{p q}^r $\\[2mm]
$\bar{b}{}_{a p }$ & $b{}_{b q }$ & ${G}_{\mu r }$ & 
 & $ g_s\gamma_{a b}^\mu \lambda_{p q}^r $\\[2mm]
$\bar{u}{}_{a p }$ & $s{}_{b q }$ & $W^+_F{}_{}$ &
  & $\frac{1}{4}\frac{ i e M_s\sqrt{2} Vus}{ MW
 s_w}\delta_{p q} \big(\delta_{a b} +\gamma_{a b}^5 \big)$\\[2mm]
$\bar{s}{}_{a p }$ & $s{}_{b q }$ & ${H}_{}$ & 
 & $-\frac{1}{2}\frac{ e M_s}{ MW s_w}\delta_{a b} \delta_{p q} $\\[2mm]
$\bar{s}{}_{a p }$ & $s{}_{b q }$ & $Z_F{}_{}$ &
  & $-\frac{1}{2}\frac{ i e M_s}{ MW s_w}\delta_{p
 q} \gamma_{a b}^5 $\\[2mm]
$\bar{s}{}_{a p }$ & $u{}_{b q }$ & $W^-_F{}_{}$ &
  & $-\frac{1}{4}\frac{ i e M_s\sqrt{2} Vus}{ MW
 s_w}\delta_{p q} \big(\delta_{a b} -\gamma_{a b}^5 \big)$\\[2mm]
$\bar{u}{}_{a p }$ & $b{}_{b q }$ & $W^+_F{}_{}$ & 
 & $\frac{1}{4}\frac{ i e M_b\sqrt{2} Vub}{ MW
 s_w}\delta_{p q} \big(\delta_{a b} +\gamma_{a b}^5 \big)$\\[2mm]
$\bar{b}{}_{a p }$ & $b{}_{b q }$ & ${H}_{}$ & 
 & $-\frac{1}{2}\frac{ e M_b}{ MW s_w}\delta_{a b} \delta_{p q} $\\[2mm]
$\bar{b}{}_{a p }$ & $b{}_{b q }$ & $Z_F{}_{}$ &
  & $-\frac{1}{2}\frac{ i e M_b}{ MW s_w}\delta_{p
 q} \gamma_{a b}^5 $\\[2mm]
$\bar{b}{}_{a p }$ & $u{}_{b q }$ & $W^-_F{}_{}$ & 
 & $-\frac{1}{4}\frac{ i e M_b\sqrt{2} Vub}{ MW s_w}\delta_{p
 q} \big(\delta_{a b} -\gamma_{a b}^5 \big)$\\[2mm]
$\bar{c}{}_{a p }$ & $s{}_{b q }$ & $W^+_F{}_{}$ &
  & $\frac{1}{4}\frac{ i e\sqrt{2} Vcs}{ MW s_w}\delta_{p
 q} \big( M_s\delta_{a b} + M_s\gamma_{a b}^5 - M_c\delta_{a
 b} + M_c\gamma_{a b}^5 \big)$\\[2mm]
$\bar{s}{}_{a p }$ & $c{}_{b q }$ & $W^-_F{}_{}$ & 
 & $-\frac{1}{4}\frac{ i e\sqrt{2} Vcs}{ MW s_w}\delta_{p
 q} \big( M_s\delta_{a b} - M_s\gamma_{a b}^5 - M_c\delta_{a
 b} - M_c\gamma_{a b}^5 \big)$\\[2mm]
$\bar{c}{}_{a p }$ & $b{}_{b q }$ & $W^+_F{}_{}$ & 
 & $\frac{1}{4}\frac{ i e\sqrt{2} Vcb}{ MW s_w}\delta_{p
 q} \big( M_b\delta_{a b} + M_b\gamma_{a b}^5 - M_c\delta_{a
 b} + M_c\gamma_{a b}^5 \big)$\\[2mm]
$\bar{b}{}_{a p }$ & $c{}_{b q }$ & $W^-_F{}_{}$ & 
 & $-\frac{1}{4}\frac{ i e\sqrt{2} Vcb}{ MW s_w}\delta_{p
 q} \big( M_b\delta_{a b} - M_b\gamma_{a b}^5 - M_c\delta_{a
 b} - M_c\gamma_{a b}^5 \big)$\\[2mm]
$\bar{t}{}_{a p }$ & $s{}_{b q }$ & $W^+_F{}_{}$ &
  & $\frac{1}{4}\frac{ i e\sqrt{2} Vts}{ MW s_w}\delta_{p q}
 \big( M_s\delta_{a b} + M_s\gamma_{a b}^5 - M_t\delta_{a b}
 + M_t\gamma_{a b}^5 \big)$\\ \hline
\end{tabular}

\begin{tabular}{|llll|l|} \hline
\multicolumn{4}{|c|}{Fields in the vertex} &
 Variational derivative of Lagrangian by fields \\ \hline
$\bar{s}{}_{a p }$ & $t{}_{b q }$ & $W^-_F{}_{}$
 &  & $-\frac{1}{4}\frac{ i e\sqrt{2} Vts}{ MW
 s_w}\delta_{p q} \big( M_s\delta_{a b} -
 M_s\gamma_{a b}^5 - M_t\delta_{a b} - M_t\gamma_{a b}^5 \big)$\\[2mm]
$\bar{t}{}_{a p }$ & $b{}_{b q }$ & $W^+_F{}_{}$ &
  & $\frac{1}{4}\frac{ i e\sqrt{2} Vtb}{ MW s_w}\delta_{p q}
 \big( M_b\delta_{a b} + M_b\gamma_{a b}^5 - M_t\delta_{a b}
 + M_t\gamma_{a b}^5 \big)$\\[2mm]
$\bar{b}{}_{a p }$ & $t{}_{b q }$ & $W^-_F{}_{}$ & 
 & $-\frac{1}{4}\frac{ i e\sqrt{2} Vtb}{ MW s_w}\delta_{p q}
 \big( M_b\delta_{a b} - M_b\gamma_{a b}^5 - M_t\delta_{a b}
 - M_t\gamma_{a b}^5 \big)$\\[2mm]
$\bar{c}{}_{a p }$ & $c{}_{b q }$ & ${H}_{}$ &
  & $-\frac{1}{2}\frac{ e M_c}{ MW s_w}\delta_{a b} \delta_{p q} $\\[2mm]
$\bar{c}{}_{a p }$ & $c{}_{b q }$ & $Z_F{}_{}$ &  &
 $\frac{1}{2}\frac{ i e M_c}{ MW s_w}\delta_{p q} \gamma_{a b}^5 $\\[2mm]
$\bar{d}{}_{a p }$ & $c{}_{b q }$ & $W^-_F{}_{}$ &
  & $\frac{1}{4}\frac{ i e M_c\sqrt{2} Vcd}{ MW s_w}\delta_{p q}
 \big(\delta_{a b} +\gamma_{a b}^5 \big)$\\[2mm]
$\bar{c}{}_{a p }$ & $d{}_{b q }$ & $W^+_F{}_{}$ & 
 & $-\frac{1}{4}\frac{ i e M_c\sqrt{2} Vcd}{ MW
 s_w}\delta_{p q} \big(\delta_{a b} -\gamma_{a b}^5 \big)$\\[2mm]
$\bar{t}{}_{a p }$ & $t{}_{b q }$ & ${H}_{}$ & 
 & $-\frac{1}{2}\frac{ e M_t}{ MW s_w}\delta_{a b} \delta_{p q} $\\[2mm]
$\bar{t}{}_{a p }$ & $t{}_{b q }$ & $Z_F{}_{}$ &
  & $\frac{1}{2}\frac{ i e M_t}{ MW s_w}\delta_{p
 q} \gamma_{a b}^5 $\\[2mm]
$\bar{d}{}_{a p }$ & $t{}_{b q }$ & $W^-_F{}_{}$ &
  & $\frac{1}{4}\frac{ i e M_t\sqrt{2} Vtd}{ MW
 s_w}\delta_{p q} \big(\delta_{a b} +\gamma_{a b}^5 \big)$\\[2mm]
$\bar{t}{}_{a p }$ & $d{}_{b q }$ & $W^+_F{}_{}$ &
  & $-\frac{1}{4}\frac{ i e M_t\sqrt{2} Vtd}{ MW s_w}\delta_{p q}
 \big(\delta_{a b} -\gamma_{a b}^5 \big)$\\[2mm]
$\bar{\nu}^\mu{}_{a }$ & $\mu{}_{b }$ & $W^+_F{}_{}$ & 
 & $\frac{1}{4}\frac{ i e M_\mu\sqrt{2}}{ MW s_w}\big(\delta_{a
 b} +\gamma_{a b}^5 \big)$\\[2mm]
$\bar{\mu}{}_{a }$ & $\mu{}_{b }$ & ${H}_{}$ &  & $-\frac{1}{2}\frac{
 e M_\mu}{ MW s_w}\delta_{a b} $\\[2mm]
$\bar{\mu}{}_{a }$ & $\mu{}_{b }$ & $Z_F{}_{}$ &  & $-\frac{1}{2}\frac{ 
i e M_\mu}{ MW s_w}\gamma_{a b}^5 $\\[2mm]
$\bar{\mu}{}_{a }$ & $\nu^\mu{}_{b }$ & $W^-_F{}_{}$ &
  & $-\frac{1}{4}\frac{ i e M_\mu\sqrt{2}}{ MW s_w}\big(\delta_{a b}
 -\gamma_{a b}^5 \big)$\\[2mm]
$\bar{\nu}^\tau{}_{a }$ & $\tau{}_{b }$ & $W^+_F{}_{}$ &
  & $\frac{1}{4}\frac{ i e M_\tau\sqrt{2}}{ MW s_w}\big(\delta_{a b}
 +\gamma_{a b}^5 \big)$\\[2mm]
$\bar{\tau}{}_{a }$ & $\tau{}_{b }$ & ${H}_{}$ &  & $-\frac{1}{2}\frac{
 e M_\tau}{ MW s_w}\delta_{a b} $\\[2mm]
$\bar{\tau}{}_{a }$ & $\tau{}_{b }$ & $Z_F{}_{}$ &  & $-\frac{1}{2}\frac{
 i e M_\tau}{ MW s_w}\gamma_{a b}^5 $\\[2mm]
$\bar{\tau}{}_{a }$ & $\nu^\tau{}_{b }$ & $W^-_F{}_{}$ & 
 & $-\frac{1}{4}\frac{ i e M_\tau\sqrt{2}}{ MW s_w}\big(\delta_{a b}
 -\gamma_{a b}^5 \big)$\\[2mm]
$W^+_F{}_{}$ & $W^+_F{}_{}$ & $W^-_F{}_{}$ & $W^-_F{}_{}$ &
 $-\frac{1}{2}\frac{ e^2  MH^2 }{ MW^2  s_w^2 }$\\[2mm]
${H}_{}$ & $W^+_F{}_{}$ & $W^-_F{}_{}$ & 
 & $-\frac{1}{2}\frac{ e MH^2 }{ MW s_w}$\\[2mm]
${H}_{}$ & ${H}_{}$ & $W^+_F{}_{}$ & $W^-_F{}_{}$ &
 $-\frac{1}{4}\frac{ e^2  MH^2 }{ MW^2  s_w^2 }$\\[2mm]
$W^+_F{}_{}$ & $W^-_F{}_{}$ & $Z_F{}_{}$ & $Z_F{}_{}$
 & $-\frac{1}{4}\frac{ e^2  MH^2 }{ MW^2  s_w^2 }$\\[2mm]
${H}_{}$ & ${H}_{}$ & ${H}_{}$ &  & $-\frac{3}{2}\frac{ e
 MH^2 }{ MW s_w}$\\[2mm]
${H}_{}$ & $Z_F{}_{}$ & $Z_F{}_{}$ &  & $-\frac{1}{2}\frac{ 
e MH^2 }{ MW s_w}$\\[2mm]
${H}_{}$ & ${H}_{}$ & ${H}_{}$ & ${H}_{}$ & $-\frac{3}{4}\frac{
 e^2  MH^2 }{ MW^2  s_w^2 }$\\[2mm]
${H}_{}$ & ${H}_{}$ & $Z_F{}_{}$ & $Z_F{}_{}$ & $-\frac{1}{4}\frac{
 e^2  MH^2 }{ MW^2  s_w^2 }$\\[2mm]
$Z_F{}_{}$ & $Z_F{}_{}$ & $Z_F{}_{}$ & $Z_F{}_{}$ & $-\frac{3}{4}
\frac{ e^2  MH^2 }{ MW^2  s_w^2 }$\\[2mm]
$W^+_F{}_{}$ & $W^-_F{}_{}$ & ${Z}_{\mu }$ &  & $-\frac{1}{2}\frac{
 e}{ c_w s_w}\big(2 s_w^2 p_2^\mu -p_2^\mu -2 s_w^2 p_1^\mu +p_1^\mu
 \big)$\\[2mm]
${H}_{}$ & ${Z}_{\mu }$ & $Z_F{}_{}$ &  & $-\frac{1}{2}\frac{ i e}{
 c_w s_w}\big(-p_3^\mu +p_1^\mu \big)$\\[2mm]
${A}_{\mu }$ & $W^+_F{}_{}$ & $W^-_F{}_{}$ &  & $ e\big(p_3^\mu
 -p_2^\mu \big)$\\[2mm]
${H}_{}$ & $W^+_F{}_{}$ & $W^-{}_{\mu }$ &  & $-\frac{1}{2}\frac{
 i e}{ s_w}\big(p_1^\mu -p_2^\mu \big)$\\[2mm]
$W^+_F{}_{}$ & $W^-{}_{\mu }$ & $Z_F{}_{}$ &  & $-\frac{1}{2}\frac{
 e}{ s_w}\big(p_3^\mu -p_1^\mu \big)$\\[2mm]
${H}_{}$ & $W^+{}_{\mu }$ & $W^-_F{}_{}$ &  & $\frac{1}{2}\frac{
 i e}{ s_w}\big(p_3^\mu -p_1^\mu \big)$\\[2mm]
$W^+{}_{\mu }$ & $W^-_F{}_{}$ & $Z_F{}_{}$ &  & $-\frac{1}{2}\frac{
 e}{ s_w}\big(p_2^\mu -p_3^\mu \big)$\\[2mm]
$W^+_F{}_{}$ & $W^-_F{}_{}$ & ${Z}_{\mu }$ & ${Z}_{\nu }$ & $\frac{
1}{2}\frac{ e^2 }{ c_w^2  s_w^2 }g^{\mu \nu} \big(4 s_w^4 -4 s_w^2
 +\big)$\\ \hline
\end{tabular}

\begin{tabular}{|llll|l|} \hline
\multicolumn{4}{|c|}{Fields in the vertex} & Variational derivative
 of Lagrangian by fields \\ \hline
${H}_{}$ & ${Z}_{\mu }$ & ${Z}_{\nu }$ &  & $\frac{ e MW}{ c_w^2 
 s_w}g^{\mu \nu} $\\[2mm]
${H}_{}$ & ${H}_{}$ & ${Z}_{\mu }$ & ${Z}_{\nu }$ & $\frac{1}{2}\frac{
 e^2 }{ c_w^2  s_w^2 }g^{\mu \nu} $\\[2mm]
${Z}_{\mu }$ & ${Z}_{\nu }$ & $Z_F{}_{}$ & $Z_F{}_{}$ & $\frac{1}{2}
\frac{ e^2 }{ c_w^2  s_w^2 }g^{\mu \nu} $\\[2mm]
${A}_{\mu }$ & $W^+_F{}_{}$ & $W^-_F{}_{}$ & ${Z}_{\nu }$ & $-\frac{
 e^2 }{ c_w s_w}g^{\mu \nu} \big(2 s_w^2 -\big)$\\[2mm]
$W^+_F{}_{}$ & $W^-{}_{\mu }$ & ${Z}_{\nu }$ &  & $\frac{ i e MW s_w}{
 c_w}g^{\mu \nu} $\\[2mm]
${H}_{}$ & $W^+_F{}_{}$ & $W^-{}_{\mu }$ & ${Z}_{\nu }$ & $\frac{1}{2}
\frac{ i e^2 }{ c_w}g^{\mu \nu} $\\[2mm]
$W^+_F{}_{}$ & $W^-{}_{\mu }$ & ${Z}_{\nu }$ & $Z_F{}_{}$ & $\frac{1}{2}
\frac{ e^2 }{ c_w}g^{\mu \nu} $\\[2mm]
$W^+{}_{\mu }$ & $W^-_F{}_{}$ & ${Z}_{\nu }$ & 
 & $-\frac{ i e MW s_w}{ c_w}g^{\mu \nu} $\\[2mm]
${H}_{}$ & $W^+{}_{\mu }$ & $W^-_F{}_{}$ & ${Z}_{\nu
 }$ & $-\frac{1}{2}\frac{ i e^2 }{ c_w}g^{\mu \nu} $\\[2mm]
$W^+{}_{\mu }$ & $W^-_F{}_{}$ & ${Z}_{\nu }$ & $Z_F{}_{}$ &
 $\frac{1}{2}\frac{ e^2 }{ c_w}g^{\mu \nu} $\\[2mm]
${A}_{\mu }$ & ${A}_{\nu }$ & $W^+_F{}_{}$ & $W^-_F{}_{}$ &
 $2 e^2 g^{\mu \nu} $\\[2mm]
${A}_{\mu }$ & $W^+_F{}_{}$ & $W^-{}_{\nu }$ &  & $- i e
 MWg^{\mu \nu} $\\[2mm]
${A}_{\mu }$ & ${H}_{}$ & $W^+_F{}_{}$ & $W^-{}_{\nu }$ &
 $-\frac{1}{2}\frac{ i e^2 }{ s_w}g^{\mu \nu} $\\[2mm]
${A}_{\mu }$ & $W^+_F{}_{}$ & $W^-{}_{\nu }$ & $Z_F{}_{}$
 & $-\frac{1}{2}\frac{ e^2 }{ s_w}g^{\mu \nu} $\\[2mm]
${A}_{\mu }$ & $W^+{}_{\nu }$ & $W^-_F{}_{}$ &  & $ i e
 MWg^{\mu \nu} $\\[2mm]
${A}_{\mu }$ & ${H}_{}$ & $W^+{}_{\nu }$ & $W^-_F{}_{}$ &
 $\frac{1}{2}\frac{ i e^2 }{ s_w}g^{\mu \nu} $\\[2mm]
${A}_{\mu }$ & $W^+{}_{\nu }$ & $W^-_F{}_{}$ & $Z_F{}_{}$ &
 $-\frac{1}{2}\frac{ e^2 }{ s_w}g^{\mu \nu} $\\[2mm]
$W^+{}_{\mu }$ & $W^+_F{}_{}$ & $W^-{}_{\nu }$ & $W^-_F{}_{}$ &
 $\frac{1}{2}\frac{ e^2 }{ s_w^2 }g^{\mu \nu} $\\[2mm]
${H}_{}$ & $W^+{}_{\mu }$ & $W^-{}_{\nu }$ &  & $\frac{ e MW}{
 s_w}g^{\mu \nu} $\\[2mm]
${H}_{}$ & ${H}_{}$ & $W^+{}_{\mu }$ & $W^-{}_{\nu }$ & 
$\frac{1}{2}\frac{ e^2 }{ s_w^2 }g^{\mu \nu} $\\[2mm]
$W^+{}_{\mu }$ & $W^-{}_{\nu }$ & $Z_F{}_{}$ & $Z_F{}_{}$ &
 $\frac{1}{2}\frac{ e^2 }{ s_w^2 }g^{\mu \nu} $\\[2mm]
${G}_{\mu p }$ & $\bar{C}^G{}_{q }$ & $C^G{}_{r }$ &  &
 $ g_sp_3^\mu f_{p q r} $\\[2mm]
$C^{W+}{}_{}$ & $W^-{}_{\mu }$ & $\bar{C}^{Z}{}_{}$ &  &
 $\frac{ c_w e}{ s_w}p_3^\mu $\\[2mm]
$\bar{C}^A{}_{}$ & $C^{W+}{}_{}$ & $W^-{}_{\mu }$ &  &
 $ ep_1^\mu $\\[2mm]
$C^{W+}{}_{}$ & $\bar{C}^{W-}{}_{}$ & ${Z}_{\mu }$ &  &
 $-\frac{ c_w e}{ s_w}p_2^\mu $\\[2mm]
${A}_{\mu }$ & $C^{W+}{}_{}$ & $\bar{C}^{W-}{}_{}$ &  &
 $- ep_3^\mu $\\[2mm]
$W^+{}_{\mu }$ & $C^{W-}{}_{}$ & $\bar{C}^{Z}{}_{}$ &  &
 $-\frac{ c_w e}{ s_w}p_3^\mu $\\[2mm]
$\bar{C}^A{}_{}$ & $W^+{}_{\mu }$ & $C^{W-}{}_{}$ &  & 
$- ep_1^\mu $\\[2mm]
$\bar{C}^{W+}{}_{}$ & $C^{W-}{}_{}$ & ${Z}_{\mu }$ &  &
 $\frac{ c_w e}{ s_w}p_1^\mu $\\[2mm]
${A}_{\mu }$ & $\bar{C}^{W+}{}_{}$ & $C^{W-}{}_{}$ &  &
 $ ep_2^\mu $\\[2mm]
$W^+{}_{\mu }$ & $\bar{C}^{W-}{}_{}$ & $C^{Z}{}_{}$ &  &
 $\frac{ c_w e}{ s_w}p_2^\mu $\\[2mm]
$\bar{C}^{W+}{}_{}$ & $W^-{}_{\mu }$ & $C^{Z}{}_{}$ &  &
 $-\frac{ c_w e}{ s_w}p_1^\mu $\\[2mm]
$C^{Z}{}_{}$ & $W^+{}_{\mu }$ & $\bar{C}^{W-}{}_{}$ &  &
 $ ep_3^\mu $\\[2mm]
$C^{Z}{}_{}$ & $\bar{C}^{W+}{}_{}$ & $W^-{}_{\mu }$ &  &
 $- ep_2^\mu $\\[2mm]
${H}_{}$ & $C^{W+}{}_{}$ & $\bar{C}^{W-}{}_{}$ &  &
 $-\frac{1}{2}\frac{ e MW}{ s_w}$\\[2mm]
${H}_{}$ & $\bar{C}^{W+}{}_{}$ & $C^{W-}{}_{}$ &  &
 $-\frac{1}{2}\frac{ e MW}{ s_w}$\\ \hline
\end{tabular}

\begin{tabular}{|llll|l|} \hline
\multicolumn{4}{|c|}{Fields in the vertex} & 
Variational derivative of Lagrangian by fields \\ \hline
$C^{W+}{}_{}$ & $\bar{C}^{W-}{}_{}$ & $Z_F{}_{}$ & 
 & $-\frac{1}{2}\frac{ i e MW}{ s_w}$\\[2mm]
$\bar{C}^{W+}{}_{}$ & $C^{W-}{}_{}$ & $Z_F{}_{}$ & 
 & $\frac{1}{2}\frac{ i e MW}{ s_w}$\\[2mm]
${H}_{}$ & $\bar{C}^{Z}{}_{}$ & $C^{Z}{}_{}$ & 
 & $-\frac{1}{2}\frac{ e MW}{ c_w^2  s_w}$\\[2mm]
$W^+_F{}_{}$ & $\bar{C}^{W-}{}_{}$ & $C^{Z}{}_{}$ &
  & $\frac{1}{2}\frac{ i e MW}{ c_w s_w}\big(-2 s_w^2 \big)$\\[2mm]
$W^+_F{}_{}$ & $C^{W-}{}_{}$ & $\bar{C}^{Z}{}_{}$ & 
 & $\frac{1}{2}\frac{ i e MW}{ c_w s_w}$\\[2mm]
$C^{Z}{}_{}$ & $W^+_F{}_{}$ & $\bar{C}^{W-}{}_{}$ &  & $ i e MW$\\[2mm]
$\bar{C}^{W+}{}_{}$ & $W^-_F{}_{}$ & $C^{Z}{}_{}$ & 
 & $-\frac{1}{2}\frac{ i e MW}{ c_w s_w}\big(-2 s_w^2 \big)$\\[2mm]
$C^{W+}{}_{}$ & $W^-_F{}_{}$ & $\bar{C}^{Z}{}_{}$ & 
 & $-\frac{1}{2}\frac{ i e MW}{ c_w s_w}$\\[2mm]
$C^{Z}{}_{}$ & $\bar{C}^{W+}{}_{}$ & $W^-_F{}_{}$ &  & $- i e MW$\\ \hline
\end{tabular}
\end{center}

\vskip 3cm

\section*{Acknowlegements}

This work was supported by ISSEP grant a96-401, INTAS grant 93-1180-ext,
 and by the Grant Center
for Natural Sciences of State Committee for Higher Education of Russia.

\eject

\tableofcontents

\end{document}